\documentclass[epsf,aps,twocolumn,nofootinbib]{revtex4}
\usepackage{amssymb}

\usepackage{graphicx}
\usepackage{amsmath}
\usepackage{epsfig}

\def\ba{\begin{eqnarray}}
\def\ea{\end{eqnarray}}
\def\be{\begin{equation}}
\def\ee{\end{equation}}

\def\C{\mathcal{C}}
\def\L{\mathcal{L}}
\def\K{\mathcal{K}}
\def\F{\mathcal{F}}
\def\dd{\left|\partial d\right|}
\def\del{\nabla}
\def\Et{\tilde{E}}
\def\Bt{\tilde{B}}
\def\nn{\nonumber}
\def\cosech{\mathrm{cosech}}
\def\cosec{\mathrm{cosec}}
\def\rfive{^{(5)}\! R}
\def\x{\mathbf{x}}
\def\d{\mathrm{d}}

\begin{document}


\title{High-energy theory for close Randall Sundrum branes}

\author{Claudia de Rham{\footnote{e-mail address:
      C.deRham@damtp.cam.ac.uk}}\  and\
Samuel Webster{\footnote{e-mail address: S.L.Webster@damtp.cam.ac.uk}}\vspace{5pt}}

\affiliation{Department of Applied
Mathematics and Theoretical Physics\\
University of Cambridge \\
Wilberforce Road, Cambridge CB3 0WA, England}

DAMTP-2005-41

%
\begin{abstract}
We obtain an effective theory for the radion dynamics of the two-brane
Randall Sundrum model, correct to all orders in brane velocity in the
limit of close separation, which is of interest for studying brane
collisions and early Universe cosmology. Obtained via a recursive
solution of the Bulk equation of motions, the resulting theory
 represents a simple extension of the corresponding low-energy
effective theory to the high energy regime.
The four-dimensional low-energy theory is indeed not valid when corrections at second
order in velocity are considered. This extension has the remarkable
property of including only second derivatives and powers of first
order derivatives.
This important feature makes the theory particularly easy to solve. We
then extend the theory by introducing a potential and detuning the
branes.  \vspace{45pt}
\end{abstract}
\maketitle
%
\section{Introduction}
Motivated by advances in string theory (in particular heterotic
M-theory), there has recently been considerable interest in models
where spacetime is effectively five-dimensional. In these models,
matter fields are confined to three-branes, membrane-like surfaces
embedded in the higher dimensional space, while gravity (and other
bulk fields) can propagate in the whole of spacetime
\cite{Gibbons:1986wg}. The cosmological consequences of these
scenarios have been widely studied (for recent reviews see
\cite{Brax:2004xh,Davis:2005au,Langlois:2002bb}). The simplest such model is that
of Randall and Sundrum \cite{Randall:1999ee,Randall:1999vf} where the
bulk is assumed to be empty apart from a cosmological constant. This
is a toy model through which braneworld ideas can be tested and,
despite its simplicity, is rich in new physics.

In the low-energy limit,
an effective four-dimensional theory can be derived on the branes
\cite{Mendes:2000wu,Khoury:2002jq,Kanno:2002ia,Shiromizu:2002qr}.
This theory predicts that near the collision, the Hubble constant on each
brane is related to the proper contraction or expansion velocity of
the fifth dimension $\dot d$ by $H^2_\pm=\frac{\dot{d}^2}{4L^2}$,
where $L$ is the Anti-de Sitter (AdS) radius related to the bulk cosmological
constant. However, an exact calculation gives the result
$H^2_\pm=\frac{1}{L^2} \tanh ^2 \left(\frac{\dot{d}
}{2}\right)$. As expected, this agrees with the four-dimensional effective
theory only
at low velocities. To lowest order in velocity, the low-energy limit
gives an accurate result for the brane geometries.
The aim of this work is to go beyond the low-energy limit and
 to develop a covariant formalism
which describes exactly these velocity corrections in the small
distance limit $d \ll L$.

Braneworld cosmology offer a entire new set of
possibilities for the production of the large scale structure and
in particular it allows scenarios with high energies or for which the
branes are moving apart at velocities which could be large
\cite{Maartens:1999hf,Langlois:2000ns,Copeland:2000hn,
Liddle:2001yq,Sami:2003my,Maartens:2003tw,Koyama:2004ap,
deRham:2004yt,Calcagni:2003sg,Calcagni:2004bh,Papantonopoulos:2004bm,
Kunze:2003vp,Liddle:2003gw}.
We may therefore address the issue of the importance of the
high-order terms in these regimes. In scenarios such as steep
inflation \cite{Copeland:2000hn,Sami:2003my},
the energy scales are important and facilitate the end of inflation.
In the Cyclic Universe \cite{Khoury:2001wf,Khoury:2001bz} as another example, it is
suggested that terms of fourth order in velocities should be
considered in order to obtain a scale invariant spectrum after the
bounce \cite{Tolley:2003nx}.
To work with such models, it seems important to understand the behaviour of these
high-energy corrections
and in particular to understand their consequences for the
production of the large scale structure, as much in the context of an expanding
as in a contracting Universe.

However, in order find a solvable theory when the low-energy
constraint is relaxed, we need to work in another limit.
The special
limit in which  we choose to work instead is the close brane
limit where the distance $d$ between the branes is small compared with the AdS radius $L$: $d\ll
L$.
 This limit is interesting in cosmology for two main reasons.
The first one is that there has been considerable interest in the
interpretation of the Big Bang as a brane collision
\cite{Khoury:2001wf,Khoury:2001bz,Tolley:2003nx,Webster:2004sp,Gibbons:2005rt,Jones:2002cv,Creminelli:2004jg}.
Therefore, at the
beginning of our universe, there will be a regime where the distance between the branes
is very small. Such a limit seems therefore to be relevant to study the production of the large scale
structure.\\
The second reason is that it is in the limit where the distance between the
branes is small that we expect the geometry on the brane to be
well-described by a four-dimensional theory. Indeed, when the branes are
far apart, the bulk degrees of freedom (leading to the Kaluza-Klein corrections on the
branes) are more easily excited, making the theory on the branes
non-local \cite{Mukohyama:2000ui,Wiseman:2002nn,Seahra:2005us,deRham:2004yt}.

Following these arguments, in the small
distance limit, we therefore expect the branes to be well described
by a four-dimensional theory which will be valid for any
energy scale and will include higher-order derivatives. \\
To work in this regime, we follow part of the idea of Shiromizu
{\it et.al.} in \cite{Shiromizu:2002ve} although our final result will be different. The
main idea of this paper is to express the extrinsic curvature on the
negative-tension brane as a Taylor expansion in terms of the extrinsic curvature on the
positive-tension brane and its derivatives along the normal direction.
 We can then use the five-dimensional Einstein equations to express
any second (or higher) order normal derivatives of the extrinsic curvature in
terms of first normal derivatives and derivatives along the four transverse
directions.
Since the normal derivative of the extrinsic
curvature on the positive-tension brane is known up to the induced Weyl
tensor on that brane, this gives a formal equation for the induced Weyl tensor on the positive-tension
brane, which is the only unknown information on the brane.

Although formally correct, this process can be complicated, in
particular if matter is introduced on the branes.  In order to
understand this expansion we will therefore focus on the case of empty
branes for which the extrinsic curvature on the branes is
proportional to the metric. As it will appear later, the formalism
becomes very simple in that context. As already mentioned, the second condition we impose is the close
brane regime. More concretely, by imposing this condition we will keep
only the terms at leading orders in the
distance $d$ between the branes at each order in the Taylor expansion.
To do this, we will work in a recursive way.
This will enable us to work out the exact form for the
Weyl tensor on the positive-tension brane to leading order in $d$.

As we might expect the exact expression for the Weyl tensor in the
close brane limit has some higher derivative correction terms
which are not present in the low-energy four-dimensional effective
theory. However we may check that, at low velocities, this expression for
the Weyl tensor is consistent with the one derived from the
low-energy effective theory  in the small $d$ limit. Furthermore, as a
non trivial check we may verify that our result gives precisely the
right result to leading order in $d$ for the background solution where
it is possible to solve the
five-dimensional
geometry exactly. However the prescription is completely covariant
and will allow us to study perturbations without needing to solve
the full five-dimensional equations.

Having an exact expression for the induced Weyl tensor in the
close-brane limit provides us with a unique modified Einstein equation on the
brane describing the coupling of a scalar field (the radion) to the brane geometry.
The equation of motion of the scalar field is then given by the
requirement that the Weyl tensor be traceless. One of the most
promising routes is the possibility of interpreting this scalar
field as playing the role of the inflaton scalar field on the
positive-tension brane 
\cite{Dvali:1998pa,Himemoto:2000nd,Brandenberger:2003py,Quevedo:2002xw,Garousi:2004uf,Hsu:2003cy}.
During inflation both branes could be moving
apart. When the negative-tension brane moves towards infinity, its
effect on the positive-tension brane becomes negligible which means
that the scalar field decouples, giving an explanation of why such a field is not
seen on the brane at the present time. In order to interpret the
radion as a candidate for the inflaton a potential for the radion
has to be introduced. So far the exact origin of that potential is
not fully understood, but for the purpose of this study we shall
introduce a potential by hand assuming some five-dimensional effects
might explain its presence. We will therefore study how the theory
we derived on the brane can be modified in order to include such a
potential and how the equation of motion for the scalar field has to
be modified in a consistent way in order to respect the conservation
of energy and to be consistent with
the low-energy theory. In order to do so we will review how the
introduction of the potential is usually performed in the low-energy
limit and then proceed in a similar way for the close-brane theory.

The paper is organised as follows. In Section \ref{Background}, we
review the Randall Sundrum model and derive the exact Friedmann
equations on the branes for the background. We then give an overview
of the low-energy four-dimensional effective theory and compare its
predictions in the background with the exact solution. Since they
disagree beyond the leading term in velocity, in Section
\ref{cov small d}, we derive a covariant
expression for the Weyl tensor on each brane in the close
brane limit. First we start with a toy model in order to understand the
procedure. Then we present the five-dimensional model and derive the Weyl tensor on
each brane using the small $d$ approximation $d \ll L$. We use this
result in Section \ref{4d theory small d} in order
to find the exact theory on the branes, valid in the close brane
limit. We check that this theory is consistent with the conservation
of energy. Finally in Section \ref{extension} we consider
extension of this
theory, first by introducing a potential and then by detuning the brane
tensions.
 We will then discuss the
implications of our results in Section \ref{conclusion}.

In Appendix \ref{appendix n case}, we present the technical details
that allows us to derive the normal derivative of the extrinsic
curvature in order to obtain the Weyl tensor on each brane. First we
proceed in a special gauge where we suppose that the $g_{yy}$
component of the metric can be taken to be independent of the fifth
dimension. Then, we show that the same
procedure is valid when this condition is relaxed and we obtain the same result.

In Appendix \ref{appendix div T}, we present the details of the
derivation of the divergence of the stress-energy tensor and show
how the equation of motion for the scalar field has to be modified
when a potential is introduced.

Finally, for completeness, we derive the evolution equation for
the Weyl tensor in Appendix \ref{appendix weyl}.

%
%
\section{Background behaviour}
\label{Background}
In what follows, we shall be interested in the two
brane Randall Sundrum model as a specific simple example of
braneworld cosmologies. In that model,
the spacetime is five dimensional, with a compact extra dimension
having the topology of an $S_1 / \mathbb{Z}_2$ orbifold. The stress
energy of the bulk is assumed to be from a pure negative
cosmological constant $|\Lambda |= \frac{6}{\kappa\, L^2}$,
$\kappa=\frac{1}{M^3_5} \equiv \frac{L}{M^2_4}$, with $M_n$ the
n-dimensional Planck mass. For simplicity, we set $\kappa=1$ for the
rest of this work.
 There are two boundary branes located at
the fixed points of the $\mathbb{Z}_2$ symmetry on which gauge and matter
fields are confined. Apart from where stated otherwise, we
 assume, for simplicity, that the tension on each brane are fine-tuned
to their canonical value $\lambda_{\pm}=\pm \frac{6}{
L}$ and we don't include any other kind of matter on the branes.

In this paper we use the index conventions that Greek indices are
four dimensional, labeling the transverse $x^\mu$ directions, while
Roman indices are fully five dimensional.
We will denote by a ``dot'' any derivative with respect of the proper
time unless specified otherwise.

\subsection{Five-dimensional Theory}
\label{5d}
\subsubsection{Frame where the bulk is static}
In the fine-tuned case with cosmological symmetry (i.e. the three
spatial directions are assumed to be homogeneous and isotropic) one
can make considerable exact analytical progress. We work in
the frame where the bulk is static, which exists as a consequence of
Birkhoff's theorem. In that frame, the most general geometry is
Schwarzschild-Anti-de-Sitter (SAdS) with the parameter
$\mathcal{C}$ associated with the Black Hole mass \cite{Binetruy:1999hy}:
\ba
\d s^2&=&\d Y^2-n^2(Y)\d T^2+a^2(Y)\d\x^2 \\
\text{with\ \ \ } a^2(Y)&=&e^{-2 Y/L}+\frac{\mathcal{C}}{4}\ e^{2 Y/L}\nn\\
n^2(Y)&=&L^2a'(Y)^2
=a^2-\frac{\C}{a^2},\nn
\ea
where we are assuming flat spatial sections for simplicity. In this frame,
the branes are not static but have loci $Y= Y_{\pm}(T)$.
The Isra\"el matching conditions \cite{Israel}, associated with the
$\mathbb{Z}_2$-symmetry, impose the condition on the extrinsic
curvature:
\be
K_{\mu \nu}(Y=Y_\pm)=-\frac{1}{L}g_{\mu \nu}(Y=Y_\pm).
\ee
On the brane, the spatial components of the extrinsic curvature are:
\be
K_{ij}(Y_\pm)=
\left(1-\frac{\dot{{Y}}_\pm^2}{{n}_\pm^2}\right)^{-1/2}
\ \left. \frac{a'(Y)}{a(Y)}\ g_{ij}\right|_{Y=Y\pm},
\ee
so the brane velocities must satisfy:
\be
\dot{{Y}}_{\pm}^2=\left(\frac{\d {Y}_{\pm}}{\d T}\right)^2
={n}^2_\pm\, \left(1-\frac{{n}^2_\pm}{{a}^2_\pm}
\right),\label{brane velocities}
\ee
where $ a_{\pm}(T) \equiv a(Y= Y_{\pm}(T))$ is the value of
the scale factor on each brane, and similarly for $
n_{\pm}(T)$. The induced line element on the branes can be read off as
\ba
\d s_{\pm}^2&=&-( n^2_\pm-\dot{{Y}}_{\pm}^2) \d T^2+
a_{\pm}^2 \d \x^2
\nn\\
&\equiv& a_{\pm}^2\left(-\d\tau_{\pm}^2+\d\x^2 \right),
\ea
defining the conformal time on each brane:
\be
\d \tau_\pm\equiv\sqrt{\frac{ n^2_\pm-\dot{{Y}}_{\pm}^2}{
a^2_\pm}} \d T =\frac{{n}^2_\pm}{ a^2_\pm}\d T.
\ee
In terms of
the conformal time, the scale factor on each brane evolves with
constant velocity:
\ba
\left( \frac{\d  {a}_\pm}{\d\tau_\pm} \right)^2= \dot{{  Y}}_{\pm}^2
 \,
 \left(\frac{\d T}{\d\tau_\pm}\right)^2
\, \left. \frac{\d a(Y)}{\d Y} \right|^2_{Y= {Y}_{\pm}}
 =\frac{\mathcal C}{L^2},  \label{a' exact}
\ea
leading to the Hubble parameter on each brane:
\ba
H^2_\pm&=&
\frac{1}{ {a}_{\pm}^4}\left(\frac{\d  {a}_{\pm}}{\d \tau_\pm}
\right)^2 \notag \\
&=&\frac{1}{L^2}\left(1- \frac{{n}^2_\pm}{ a^2_\pm}\right)
\label{H^2}
=\frac{\C}{L^2 {a}_{\pm}^4}.
\ea
This result 
is a
direct consequence of the projected Einstein equations on the brane
(see Section \ref{closebranes}), which, for empty Friedmann Robertson
Walkers (FRW) branes, gives
the induced Ricci scalars as
\ba
R^{(\pm)}=\frac{6}{{  a}_\pm^3}\frac{\d^2 {
    a}_\pm}{\d\tau_\pm^2}=0.
\ea
We may point out that in (\ref{H^2}), the Hubble parameter on each brane is bounded by
$L^2H^2_\pm\le1$, with equality for the negative-tension brane when
this one touches the five-dimensional Black Hole horizon. This study is beyond the purpose of this
paper and we consider both branes to be far away from the five
dimensional singularity.

We now consider the specific case where
the branes are moving apart after a collision at $\tau=0$
and we denote by $a_0$ the value of the scale factor when the branes
coincide (the situation where the branes are moving towards each other before
the collision is completely analogous).

In contrast to the exact treatment above we now only consider the
system in the limit of small brane separation, i.e. ${a}_+ \approx
{a}_- \approx a_0 $,
$\ {n}_+ \approx {n}_- \approx n_0$ and $\dot{{Y}}_+\approx
-
\dot{{Y}}_-$. Using (\ref{brane velocities}), to linear order we have:
\ba
{Y}_+(T)&=&Y_0-v (T-T_0)\left( 1+\mathcal{O}\left(\frac{T-T_0}{L}\right) \right)\\
{Y}_-(T)&=&Y_0+v (T-T_0)
\left( 1+\mathcal{O}\left(\frac{T-T_0}{L}\right) \right) \\
v&=&n_0\sqrt{1-\frac{n_0^2}{a_0^2}},\label{v}
\ea
where the collision happens at $T=T_0$ and
$Y=Y_0$. When the branes are close, the proper distance between the
branes goes as $d \sim (T-T_0)$, so the corrections are of order
$\mathcal{O}\left(\frac{T-T_0}{L}\right) =
\mathcal{O}\left(\frac{d}{L}\right)$.
This is the limit in which we will work all through this paper.

\subsubsection{Frame where the branes are quasi-static }
So far we have worked in the frame where the bulk was
static. However if we are interested in the proper distance between
the branes, it is more intuitive to derive it from the frame where the
branes are static. Such a frame is in general complicated, however, for the
purpose of this study, it is enough to consider the frame in which
the branes are ``quasi-static", or static to leading order in $d/L$
or in $(T-T_0)/L$.
In order to work in such a frame,
 we may perform the gauge transformation $\left(Y,T\right)\rightarrow \left(y,t \right)$:
\ba
T&=&T_0+\frac{t}{n_0} \cosh \left(\left(2y-1\right)\tanh^{-1}
\left(\frac{v} {n_0}\right)\right)\\
Y&=& Y_0+\, t \ \sinh \left(\left(2y-1\right)\tanh^{-1}
\left(\frac{v} {n_0}\right)\right).
\ea
We can indeed check that at $y=0$, $Y=Y_0-v (T-T_0)$
 and at $y=1$, $Y=Y_0+v(T-T_0)$.
In this new frame, the branes are static to leading order in $d$ and located at
$y=0$ and $y=1$. In terms of the new coordinates $y$ and $t$ the bulk
geometry in the limit of close separation is:
\ba
\d s^2&=& A(t)^2\, \d y^2
-\d t^2+a^2 \d\x^2, \label{bckd frame brane static} \\
\text{with}\ A(t)&=&2\, t \tanh^{-1}\left(\frac{v} {n_0}\right).
\ea
In this limit, the induced metric on the branes is then
$ \d s^2_\pm=-\d t^2+a(t)^2 \d\x^2$,  where $t$
is the proper time.
When $d \ll L$, the proper distance between the branes is given by:
\ba
d&=&\int_0^1  A(t)\,  dy \notag \\
&=&2 \, t \,
\tanh^{-1}\left(\frac{v}{n_0}\right),
\ea
so that, to leading order, the expansion of
the fifth dimension with respect to the proper time $t$ is
\be
\dot d=2\tanh^{-1}(\frac{v}{n_0})\label{d(t)},
\ee
where $v$ is given in (\ref{v}). At the collision, $v$ is related to the
Hubble parameter (\ref{H^2}) by:
\ba
\frac{v^2}{n_0^2}=1-\frac{n_0^2}{a_0^2}=L^2H^2_\pm(\tau=0),
\ea
giving the relation between the Hubble parameter and the $\dot d$:
\be
H_\pm=\pm \frac{1}{L}\tanh \left(\frac{\dot d}{2}
\right), \ \text{for } d \ll L.
\label{H exact}\ee
This result is derived directly from the full five-dimensional
equations subject only to the assumption that the branes are close.
In (\ref{H^2}), we have already pointed out that the Hubble
parameter was bounded $L^2 H_\pm^2 <1$ which is consistent with (\ref{H
exact}).

We can compare this result with the analogous relation derived from the
standard low-energy effective theory which we shall briefly review in
the next Section. In Section \ref{4d theory small d}
we will then compare this result with the one obtained from the
small-$d$ theory which we derived in this paper and we shall see that they agree perfectly.

\subsection{Four-dimensional effective theory at low energy}

\subsubsection{Low-energy effective theory}
\label{section 4d eff}
At low energies the system is well-described by a four-dimensional
effective theory \cite{Kanno:2002ia,Shiromizu:2002qr,Binetruy:2001tc}.
In this limit, braneworlds behave like conventional scalar-tensor
theory of gravity where the bulk effects are
represented by a single scalar field. In the low-energy approximation
the induced metrics on each brane are conformally related:
\begin{eqnarray}
g_{\mu \nu}^{(-)}=\Psi^2 g_{\mu \nu}^{(+)}, \label{confrel}
\end{eqnarray}
and the field $\Psi$ is related to the proper distance $d$ between
the branes by $\Psi=e^{-d/L}$. The system is closed and described by
the modified Einstein equation:
\begin{eqnarray}
G^{(+)}_{\mu \nu} \hspace{-5pt}&=& \hspace{-5pt} -E_{\mu \nu}^{(+)}\\
 \hspace{-5pt}&=& \hspace{-5pt}
 \Psi^2 G^{(+)}_{\mu \nu}+4 \partial_{\mu }\Psi
\partial_{\nu }\Psi
-2\Psi D_{\mu} D_{\nu} \Psi \label{Euv+}\\
 \hspace{-5pt}&& \hspace{-5pt}+\left(2 \Psi \Box \Psi
-\left(\partial \Psi \right)^2
\right)g_{\mu\nu}^{(+)}, \notag
\end{eqnarray}
where all covariant derivatives are taken with respect to
$g_{\mu\nu}^{(+)}$.
This is in complete agreement with the Gauss-Codacci equations in the low-energy limit.
The equation of motion for the scalar field is
given by the requirement that $E_{\mu \nu}^{(+)}$
is traceless:
\ba
\Box  \Psi=0.
\ea
In
terms of the proper distance between the branes,
the low-energy effective theory becomes:
\begin{eqnarray}
G^{(+)}_{\mu \nu}&=&
\frac{2}{L}\frac{\Psi^2}{1-\Psi^2}
\left[
D_\mu D_\nu d\right. \label{Einstein eff in d}\\
&& \left.
+\frac{1}{L}\left(\partial_{\mu} d \partial_{\nu} d-\frac{1}{2}
\left(\partial d\right)^2g_{\mu\nu}^{(+)} \right)
\right] \notag\\
\Box \, d&=&\frac{\left(\partial d\right)^2}{L}.
\label{box d eff}
\end{eqnarray}
\subsubsection{Background behaviour of the effective theory}
We may now examine the behaviour of this effective theory in the
special case of a flat FRW background in
order to compare the results with those of Section \ref{5d}.
We consider the special case where the positive-tension
brane expands. Since the Ricci scalar vanishes, the scale factor still satisfies:
\ba
&& \frac{\d^2 a_+}{\d\tau^2}=0,\\
&& a_+=v_+\tau+a_0,
\ea
where $a_0$ is the value of the scale
factor at the collision and $v_+$ is an arbitrary constant. Comparison
with the five-dimensional result (\ref{a' exact}) would identify $v_+$ as $\sqrt{C}/L$
but, considering only the four-dimensional effective theory, its value
is not determined. The Friedmann equation obtained from (\ref{Euv+})
is:
\be
\left(1-\Psi^2\right)H_+^2-2\Psi\dot{\Psi}H_+-\dot{\Psi}^2=0.\label{quadratic}
\ee
This is a quadratic equation with the two solutions:
\ba
H_+=\frac{\dot d }{L\left(1\pm \Psi^{-1} \right) }.\label{Friedmanneff}
\ea
The two possible signs correspond to the branes moving either in the
same $(-)$ or opposite $(+)$ directions, which can be seen as follows. From the
conformal relation (\ref{confrel}) (i.e. $a_-=\Psi a_+$), the Hubble
parameter on the negative-tension brane can be written in terms of
$H_+$ as:
\ba
H_-&\equiv&\frac{\dot{a}_-}{a_-}=-\frac{\dot d}{L}+H_+=\mp e^{d/L}H_+ \notag \\
&=& \mp H_+\left(1+\mathcal{O} (d/L)\right).
\ea
So a $-$ sign in (\ref{Friedmanneff}) corresponds to the situation
where $H_- \simeq H_+$ for which the branes are moving in the same
direction. In that case, near the collision,
\be
\dot d\approx -H_+d \approx -\frac{v_+}{a_0^2}d,\label{ddotsimd}
\ee
so the branes take an infinite time to collide.
Instead, we are interested in the situation where the collision happens at finite
time and the branes are moving in
opposite direction $H_- \simeq - H_+$, corresponding to a
$+$ sign in (\ref{Friedmanneff}). In this case,
\ba
H_\pm=\pm \frac{\dot{d}}{2L}, \ \ \text{for } d\ll L, \label{H eff}
\ea
which implies
\be
\label{ddotsimd0}\dot d\approx \frac{2Lv_+}{a_0^2}.
\ee
In that case we notice that $\dot d$ is of order $\left(d/L\right)^0$.
We may compare this result with (\ref{H exact}), which is correct to all
orders in $\dot{d}$ for small $d/L$. As expected, the low-energy theory
reproduces only the leading term. The effective theory is therefore only valid
to leading order in velocities (even for the background), but for any
value of $d$. We may point out that in this result both $L H_\pm$
and $\dot d$ appear to be unbounded. This is only true in this low-energy
regime for which the restrictions are very strong $L H_\pm \sim \dot d \ll 1$. As seen in (\ref{H
exact}) when $L H_\pm \sim 1$ some new restrictions
have to be imposed.

Since the low-energy effective theory only predicts the leading
order of velocity, it might be possible
to go beyond the low energy restriction and find a theory which
would be valid to all order in velocities. In order to derive such
a theory, we will work in a
regime where the branes are close to each other. In the rest of this
paper we show the existence and consistency of such a theory which
successfully reproduces (\ref{H exact}) and agrees with the
low-energy theory in the regime of small separation \emph{and}
velocity in which they are both valid.

%
%
\section{Covariant approach in Small $d$ limit}
\label{cov small d}
\subsection{Toy model \label{toymodel}}
In order to understand the procedure we will follow to work out the
theory on the brane in the close brane limit, we first examine
the following one-dimensional example.
We consider a second order differential equation:
\ba
f''(y)=U(y) f'(y)+V(y) f(y) +W(y), \label{example}
\ea
where $U, V$ and $W$ are some known function of $y$.\\
We assume that the function $f$ is known at $y=0$ and at $y=1$ and
we wish to find the value $f'(0)$ of its derivative at $y=0$. One way to
do it would be to solve the differential equation exactly with the
two boundary conditions for $f$. Once $f(y)$ is known for any $y$, we can infer
$f'(0)$. But by doing so we extracted more information that we
actually wanted; this method would be equivalent to solving the five-dimensional Einstein
equations exactly in order to obtain the induced geometry on the brane. Although this
is in theory possible it would be very hard to do. Instead we
will summarise in this example the method used by
\cite{Shiromizu:2002ve}. The idea is not to solve the
differential equation exactly but to differentiate it
in order to use it in the Taylor expansion:
\ba
f(y=1)=\sum_{n \ge 0} \frac{1}{n!} f^{(n)}(0).
\ea
By differentiating equation (\ref{example}), we can find an
expression for $f^{(n)}(y)$:
\ba
f^{(n)}(y)=U_{n-1}(y) f'(y)+V_{n-1}(y) f(y) +W_{n-1}(y),
\ea
where $U_n, V_n$ and $W_n$ may be found in a recursive way:
\begin{eqnarray*}
U_{n+1}(y)&=&U'_{n}(y)+V_n(y)+U(y)U_n(y),\\
&&U_1=U(y), U_0=1, U_{-1}=0 \\
V_{n+1}(y)&=&V'_{n}(y)+V(y)U_n(y), \\
&&V_1=V(y), V_0=0, V_{-1}=1 \\
W_{n+1}(y)&=&W'_{n}(y)+W(y)U_n(y), \\
&&W_1=W(y), W_0=0, W_{-1}=0.
\end{eqnarray*}
Using
these expressions, we may write $f'(0)$ as:
\ba
\label{fprime}
f'(0)&=&\frac{1}{\sum \frac{1}{n!}
U_{n-1}(0)}\left(f(1)\phantom{-\sum \frac{1}{n!}}\right.\\
&& \left.-\sum \frac{1}{n!}
\left.\left(V_{n-1}f-
W_{n-1}\right)\right|_{y=0}\right).\nn
\ea
Knowing $U_n, V_n$ and $W_n$ in a recursive way, we can perform
the sums and find an exact expression for $f'(0)$. This will be very
similar to the method we will use to find the induced Weyl tensor on
the brane.  Although the extrinsic curvature is known on the branes, its
normal derivative (which involves the Weyl tensor) is not. We can
derive, however, a second order differential equation for the
extrinsic curvature which allows us to calculate this derivative in the
same way as (\ref{fprime}).

However, already in this linear one
dimensional problem, the recursive relations are non-trivial. The five-dimensional
problem is even harder since, unsurprisingly, the equations are
 non linear and formidably complicated. However, if we keep only the
 leading terms in $d/L$, we find that the second order differential equation for $K_{\mu \nu}$ is
linear and, remarkably, the Taylor series corresponding to
(\ref{fprime}) becomes tractable. We therefore end up with an
expression, correct to leading order in $d/L$, for the normal
derivative of the extrinsic curvature on the positive-tension brane
which enables us to write down the Einstein equations and obtain the
close brane limit of the exact theory on the brane.

\subsection{Regime of validity}
From now on we will work in a regime where the branes are very close $d
\ll
L$.
As already mentioned, there are two type of solutions for the
background, depending whether the branes are moving in the same or in opposite directions.
If the branes are moving in the same direction, we have seen in
(\ref{ddotsimd}) that,
for the background, $\dot d \sim d/L$. It is subject to this assumption
that \cite{Shiromizu:2002ve} was derived. In this regime, we may check that
to leading order in $d$, they recover the low-energy effective theory.
Therefore, for this regime, the low-energy effective theory is valid
to all order in velocities (at small $d$). However from (\ref{ddotsimd0}) we see that
this is not valid when the branes are moving in opposite directions.
In that case,
$\dot d \sim
\left(\frac d L\right)^0\sim 1$.
This will be the solution we will be interested in for the rest
of this work and we will assume the relation $\partial_\mu d \sim
\left(\frac d L\right)^0$. Although this is strictly true only for the
background, if we work with perturbations in a comoving gauge, this relation will still
hold. In that gauge,
$\partial_\mu d \sim \left(\frac d L\right)^0$ will still be true
covariantly.

Furthermore, for adiabatic perturbations, the perturbations
behave the same way as the most general background solution. For the background
there are two kinds of solutions, one where the branes move in
the same direction for which $\partial_\mu d \sim \frac d L$ and one
where the collision happens at finite time, for which
$\partial_\mu d \sim \left(\frac d L\right)^0$.
This is true for adiabatic perturbations as
well. The adiabatic perturbations will
 follow a similar evolution to one of the two background
solutions or a superposition of them.
Since the  low-energy effective theory reproduces correctly
one type of solution we may focus on perturbations that
follow the other kind of behaviour for which
$\partial_\mu d \sim \left(\frac d L\right)^0$.

At the level of perturbations, one might think that this procedure might
break down since the perturbations diverge. But this divergence is
actually logarithmic in $d$ and is therefore negligible compared to the
terms in $1/d$ that we will find in the theory (\ref{Einstein small
d}). Compared to $L/d$, $\log(d/L)\sim(d/L)^0$.
In \cite{Creminelli:2004jg}, it is actually shown that in the
right gauge, the perturbations remain ``small'' going towards the
bounce.

\subsection{Gauss Codacci formalism in the frame where the branes are
 static}
\label{closebranes}
\subsubsection{Formalism}
In this subsection we will follow the formalism of \cite{Shiromizu:2002ve}.
We choose coordinates where the metric is of the form
\ba
ds^2&=&g_{ab}\d x^a \d x^b \nn\\
&=&A^{2}(y,x)\d y^2+q_{\mu \nu}(y,x)\d x^{\mu}\d x^{\nu},
\label{ds^2=}
\ea
with $x^\mu$ the transverse coordinates and $y$ parameterising the
extra dimension. In this frame, the branes located at the fixed positions
$y=0$ and $y=1$. $q_{\mu \nu}(\bar y, x)$
is the induced metric of a $y=\bar y=\text{const}$ hypersurface. In
particular,
$g^{(+)}_{\mu \nu}(x)=q_{\mu \nu}(y=0,
x)$ and $g^{(-)}_{\mu \nu}(x)=q_{\mu \nu}(y=1, x)$
are the induced metrics on both branes. \\
The Einstein tensor on a $y=\text{const}$ hypersurface will be written
simply as $G_{\mu\nu}$, whereas the fully five-dimensional tensor will
be denoted by $^{(5)}G_{\mu\nu}$. Four-dimensional quantities may be expressed in terms of
five-dimensional quantities by means of the Gauss-Codacci formalism
\cite{Binetruy:1999ut,Shiromizu:1999wj,Langlois:2003yy}.
Using the bulk Einstein equations
\ba
^{(5)}G_{ab}=-\frac{6}{L^2}g_{ab},
\ea
the modified four-dimensional Einstein equation is:
\ba
G_{\mu \nu}=&&\frac{3}{L^2}q_{\mu \nu} +K K_{\mu \nu}-K_{\mu
\alpha}K^{\alpha}_{\nu}\label{gauss codacci} \\
&&-\frac 1 2 \left(K^2- K^{\alpha}_{\beta}K^{\beta}_{\alpha}
\right)q_{\mu \nu} -E_{\mu \nu},\notag
\ea
where $E_{\mu \nu }$ is the Electric part of the five-dimensional
Weyl tensor (see Appendix \ref{appendix weyl}). All indices are raised and lowered with respect to
the four-dimensional metric $q_{\mu \nu}$. \\
For a $\mathbb{Z}_{2}$-symmetric brane, the extrinsic curvature $K_{\mu \nu
}$ can be uniquely determined on the brane using the Isra{\"e}l
Matching conditions, which reduce to:
\ba
K^{\mu}_{\nu}(y=0)= K^\mu_\nu(y=1)=-\frac{1}{L}\, \delta^{\mu}_{\nu}.\label{K+-}
\ea
Substituting this into (\ref{gauss codacci}) gives the projected
Einstein equation on the branes:
\be
G^{(\pm)}_{\mu\nu}=-E^{(\pm)}_{\mu\nu},
\label{proj einstein gen}
\ee
where $G^{(+)}_{\mu\nu}=G_{\mu\nu}(y=0)$ and
$G^{(-)}_{\mu\nu}=G_{\mu\nu}(y=1)$ are the induced Einstein tensors on
the two branes.
The aim of this work is
to find $E^{(\pm)}_{\mu \nu}$ exactly in the close brane approximation.
 As shown in \cite{Shiromizu:2002qr,Shiromizu:1999wj} the Weyl
 tensor can be expressed in terms of the normal derivative of the
 extrinsic curvature:
\ba
E^{\mu}_{\nu}=-\frac{1}{A}\, \partial_{y}K^{\mu}_{\nu} -\frac{D^{\mu}D_{\nu}
A}{A}-K^{\mu}_{\alpha}K^{\alpha}_{\nu}+\frac{1}{L^2}\delta^{\mu}_{\nu},
\label{Euv 1}
\ea
where $D_{\mu}$ represents covariant derivative with respect to $q_{\mu
\nu}$.
 We may define the proper distance $\tilde d$ between the positive-tension
 brane and the hypersurface located at $y=\text{const}$
by $\tilde d(y,x)=\int_0^y
 A(y',x)dy' $. The proper distance between the branes is defined as
 $d(x)=\tilde d(y=1,x)$.

The Lie derivative $\L_n\equiv \partial _{\tilde d} \equiv
\frac{1}{A}\partial_{y}$ is the derivative along the normal vector
of any $y=\text{const}$ hypersurface. In particular the extrinsic curvature
is the derivative along the normal vector of the induced metric on
such a hypersurface: $
K_{\mu \nu}(y,x)=\frac{1}{2} \partial _{\tilde d}\,  q_{\mu
\nu}(y,x)$.

In what follows we will use a similar procedure to the toy model.
Knowing the extrinsic curvature on both branes, we write an
expression for its derivative using the Taylor expansion:
\ba
K^{\mu}_{\nu}(y=1)=\sum_{n\ge 0} \frac{1}{n!}
\left. \partial _{y}^{(n)}
K^{\mu}_{\nu} \right|_{y=0}. \label{taylor1}
\ea
In what follows, we will use the
notation: $Q^{\prime} \equiv  \partial _{y} Q$ and $Q^{(n)} \equiv  \partial _{y}^{(n)} Q$ for any
quantity $Q(y,x)$ carrying indices only along the directions of the
$y=\text{const}$ hypersurface.

To start with we will consider $A$ to be independent of $y$. This is indeed the case
for the background geometry of the close-brane limit (\ref{bckd frame brane static}). We
 will see in Appendix \ref{A(y)}
 that this assumption does not affect the final
 answer. In that case $\tilde d(y,x)=y A(x)$, and in particular the
 proper distance between the branes is: $d(x)=A(x)$. We can now rewrite the
 expression (\ref{Euv 1}) in the form:
\ba
K^{\mu\, \prime}_{\ \nu}=-d E^{\mu}_{\nu}-D^{\mu}D_{\nu}
d-dK^{\mu}_{\alpha}K^{\alpha}_{\nu}+\frac{d}{L^2}\delta^{\mu}_{\nu}.
\label{Euv 2}
\ea
In order to find the expression for
$K^{ \mu \, (n)}_{\, \nu}$ in (\ref{taylor1}) we need
the derivative of the Weyl tensor and of the Christoffel Symbol. The
last one is given by:
\ba
\Gamma ^{\, \alpha\, \prime}_{\mu \nu}=
D_{\mu}(d\, K^{\alpha}_{\nu})
+D_{\nu}(d\, K^{\alpha}_{\mu}) -D^{\alpha}(d\, K_{\mu \nu}).
\label{gamma '}
\ea
For the Weyl tensor derivative, we may use the result derived in
Appendix \ref{appendix weyl}:
\ba
E^{\mu\, \prime}_{\ \nu}
&=& \hspace{-5pt}d \Big(2K^{\alpha}_{\nu} E^{\mu}_{\alpha}-\frac{3}{2} K
E^{\mu}_{\nu}-\frac 1 2 K^{\alpha}_{\beta} E^{\beta}_{\alpha}
\delta^{\mu}_{\nu} \label{E prime}\\
&& \hspace{-5pt}+C^{\mu}_{\phantom{\mu} \alpha
\nu
\beta}K^{\alpha \beta}
+(K^{3})^\mu_{\, \nu}\Big)
\notag \\ && \hspace{-5pt}
-\frac{1}{2 d}D^{\alpha} \left[
d^2 D^{\mu} K_{\alpha \nu}+d^2D_{\nu} K^{\mu}_{\alpha}-2d^2 D_{\alpha}
K^{\mu}_{\nu}
\right]\nn.
\ea
where $(K^{3})^\mu_{\, \nu}$ are some cubic terms in the
traceless part of the extrinsic curvature which exact form will not be relevant for the
purpose of this study (they vanish at $y=0$ and $y=1$). Both these
relations
(\ref{gamma '}) and (\ref{E prime}) are valid for any $y$.
On the brane they simplify considerably:
\ba
\Gamma ^{\, \alpha\, \prime}_{\mu \nu}(0)&=&
-\frac{1}{L}\left(
d_{, \mu}\delta^\alpha_\nu
+d_{, \nu}\delta^\alpha_\mu
-d^{, \alpha}g^{(+)}_{\mu \nu}
\right) \\
E^{\mu\, \prime}_{\ \nu}(0)&=&\frac{4d}{L}E^\mu_\nu(0).
\ea
The important point is that the system is now
closed. Writing $E_{\mu \nu}$ back in terms of the extrinsic
curvature and its derivative, we obtain a second order differential
equation for the extrinsic curvature which is non-linear but which
involves four-dimensional quantities only. We may therefore apply the procedure
of Section \ref{toymodel}. In order to do so we will
assume the branes to be close and keep only the leading order in $d$ in the
expansion.

\subsubsection{Small d approximation}
\label{smalldsection}
In what follow we denote by
$ ^{0}\! K^{\mu\, (n)}_{\, \nu}(x)$
the leading order in $d/L$ of $K^{\mu\, (n)}_{\, \nu}(y=0,x)$,
symbolically,
\ba
K^{(n)}=
\!\!\phantom{u}^{0}\!K^{(n)}\left(1+\mathcal{O}(d/L)\right).
\ea
Using the small distance approximation, we rewrite the expansion (\ref{taylor1}) as:
\ba
\sum_{n\ge 1} \frac{1}{n!}
 \phantom{u}^{0}\! K^{\mu\, (n)}_{\, \nu}=0, \label{taylor2}
\ea
where the $n=0$ term cancel since the extrinsic curvature
is the same on both brane (\ref{K+-}).
All the terms will be kept in the sum only if they are all of same
order. We shall see in what follows that this is indeed the case.
In order to do so, we will work in a recursive way.

\subsubsection{Example for the n=1 and n=2 case \label{n=1 and n=2}}
\label{n=1 2}
We first concentrate on the $n=1$ and $n=2$ cases in
detail in order to gain insight; the technicalities of the general $n$
case are left for Appendix \ref{appendix n case}. \\
For $n=1$,
\ba
K^{ \mu \, \prime}_{\, \nu}(y=0)=\left.-d E^{\mu}_{\nu}
- D^{\mu} D_{\nu} d\right|_{y=0}.
\label{K'(0)}
\ea
Since $\partial_{\alpha} \partial_{\beta}d \sim \partial_{\alpha}d \sim
d^0$ the second term goes as $d^{0}$.
In the effective theory, on
the brane, $E^{\mu}_{\nu} \sim d^{-1}$ (using (\ref{Einstein eff in
  d})).
Although we have argued that
at high energy the effective theory does not give the exact
expression for the Weyl tensor, we have seen that (at least for the
background) the behaviour is the same, differing only in corrections
at higher order in the velocity. In particular $E^{\mu}_{\nu}$ should go as
$d^{-1}$ at high energies as well (we will check later that
this is indeed the case).
We therefore have $K^{
\mu\, \prime}_{\, \nu}(y=0) \sim d^0$.

For the second derivative we
have:
\ba
K^{\mu\, \prime \prime}_{\, \nu}(y)=- d\, E^{\mu\, \prime }_{\, \nu}
-q^{\mu \beta \, \prime} D_{\beta} D_{\nu} d \\+
 q^{\mu \beta} \Gamma^{\, \alpha\, \prime}_{\beta \nu}
\partial_{\alpha}d-d\ \partial_{y}
\left(K^{\mu}_{\alpha}K^{\alpha}_{\nu}\right). \notag
\label{K''}
\ea
On the positive-tension brane we may compare these terms with the
ones from  $K^{\mu\, \prime }_{\,
\nu}(y=0)$:
\ba
\begin{array}{ccc}
\text{terms from $K''$} & &  \text{terms from $
K'$}\vspace{5pt}\\ 
d E^{\mu \, \prime }_{\, \nu} =\frac{4 d}{L} E^{\mu}_{\nu} & \ll &
d E^{\mu}_{\nu}, \vspace{3pt}\\
q^{\mu \beta\, \prime } D_{\beta} D_{\nu}
d=\frac{2d}{L} D^{\mu} D_{\nu} d & \ll & D^{\mu} D_{\nu} d, \vspace{3pt}\\
d \partial_{y
}\left(K^{\mu}_{\alpha}K^{\alpha}_{\nu}\right)=-\frac{2d}{L}K^{
  \mu \, \prime }_{\,
\nu} & \ll &  K^{\mu\, \prime }_{\, \nu},
\end{array}
\ea
where we used $q^{\mu \beta\, \prime }=d \partial_{\tilde d}q^{\mu
  \beta}=-2dK^{\mu \beta}$.
When $d \ll L$, the only term which is not negligible in comparison to
  the terms
present in $K^{\prime \mu}_{\ \nu}(y=0)$ is the one coming from the
derivative of the Christoffel symbol:
\ba
 q^{\mu \beta} \Gamma^{\, \alpha\, \prime}_{\beta \nu}(0)\,
\partial_{\alpha}d
&=&-\frac{1}{L}\left( 2 \partial^{\mu}d \partial_{\nu}d-
\left(\partial d \right)^2 \delta^{\mu}_{\nu}\right)\\
&\sim&
K^{\mu\, \prime }_{\, \nu}(0).\notag
\ea
This last term will give a
contribution of the same order as $K^{\mu\, \prime }_{\, \nu}(y=0)$.
%
%
\subsubsection{General n case}
In the previous specific case,
we had $ ^{0}\! K^{(1)}\sim d^{0}$ and $ ^{0}\! K^{(2)}\sim
d^{0}$.
In Appendix \ref{appendix n case}
we will show that the same is true for any $n$. In particular we
will show that the leading contribution in $K^{\mu\, (n)}_{\,
\nu}(y=0)$ comes from the $(n-1)^{\text{th}}$ derivative of this Christoffel
symbol:
\ba
  ^{0}\! K^{\mu\, (n)}_{\, \nu}=
 q^{\mu \beta} \Gamma^{\, \alpha\, (n-1)}_{\beta \nu}(0)\
\partial_{\alpha}d,
\ea
and the leading contribution from the $(n-1)^{\text{th}}$ derivative of the
Christoffel symbol comes from the term:
\ba
\Gamma^{\, \alpha\, (n-1)}_{\beta \nu}(0)&=&
d_{,\beta}  \!\! \phantom{u}^{0}\! K^{\alpha\, (n-2)}_{\, \nu}
+d_{,\nu}  \!\! \phantom{u}^{0}\! K^{\alpha\, (n-2)}_{\, \beta}\\
&&-
d^{,\alpha}\!\! \phantom{u}^{0}\! K^{(n-2)}_{\beta \nu}.\notag
\ea
Using these two results, we therefore have:
\ba
 \!\! \phantom{u}^{0}\! K^{\mu\, (n)}_{\, \nu}&=&
d^{,\mu}\!\! \phantom{u}^{0}\! K^{\alpha\, (n-2)}_{\, \nu}
+d_{,\nu}  \!\! \phantom{u}^{0}\! K^{\alpha \mu\, (n-2)}\\
&&-
d^{,\alpha}\!\! \phantom{u}^{0}\! K^{\mu\, (n-2)}_{\, \nu}.\notag
\ea
In the Taylor expansion, the term $\!\! \phantom{u}^{0}\! K^{\mu\, (n)}_{\, \nu}$
therefore comes in with the same contribution as the term
$\!\!\phantom{u}^{0}\! K^{\mu\, (n-2)}_{\, \nu}$, so all
terms have to be considered.
%
%
\subsubsection{Expression for the Weyl tensor}
We now use these results in the Taylor expansion. First,
we may define the operator $\hat O$ such that:
\ba
\hat O Z^{\mu}_{\nu}= \left[
d^{,\mu}Z^{\alpha}_{\ \nu}
+d_{,\nu} Z^{\alpha \mu}-
d^{,\alpha} Z^{\mu}_{\ \nu}
\right]d_{,\alpha},
\ea
for any four-dimensional tensor $Z^{\mu}_{\nu}$. Using this notation, the Taylor
expansion simplifies to:
\ba
\hspace{-10pt}\sum_{n\ge 0} \frac{1}{(2n+1)!}\hat O^{(n)} \!\! \phantom{u}^{0}\! K^{ \mu\, \prime}_{\, \nu}
&=&-\sum_{n\ge 1} \frac{1}{(2n)!}\hat O^{(n)}  \!\!\phantom{u}^{0}\!
K^{\mu}_{\
\nu} \label{expansion1}\\
&=&\sum_{n\ge 1} \frac{1}{L(2n)!}\hat O^{(n)} \delta^{\mu}_{\
\nu}, \notag
\ea
which may be symbolically written as:
\ba
\frac{1}{\sqrt{\hat O}} \sinh \sqrt{\hat O}\
 ^{0}\! K^{\mu\, \prime}_{\,
\nu}=\frac{1}{ L}\left(\cosh \sqrt{\hat O} -1\right)\delta^{\mu}_{\
\nu}. \label{expansion2}
\ea
To leading order in $d$, the derivative of the extrinsic
curvature on the positive-tension brane is then:
\ba
\label{operators}
K^{\mu \, \prime }_{\,
\nu}(y=0)=\left[\frac{\sqrt{\hat O}}{L}\tanh \left(\frac{\sqrt{\hat O}}{2}\right)
\right]\ \delta^{\mu}_{\
\nu}+\frac{1}{L}\mathcal{O}\hspace{-3pt}\left(\frac{d}{L}\right)
\ea
with
\ba
F^{\mu}_{\nu} \equiv \hat O \delta^{\mu}_{\nu}&=&2 \partial^{\mu}d \partial_{\nu}d -
\left(\partial d \right)^2 \delta^{\mu}_{\nu}\\
\hat O F^{\mu}_{\nu}&=& \left(\partial d \right)^4 \delta^{\mu}_{\nu}.
\ea
Therefore to leading order in $d/L$, the Weyl tensor on the positive-tension brane is:
\ba
\hspace{-10pt}E^{(+) \mu}_{\phantom{(+)}\, \nu}\hspace{-5pt}
&=&-\frac{1}{d}K^{\mu\, \prime}_{\,
\nu}(y=0)-\frac{D^{\mu}D_{\nu}d}{d},\notag\\
\label{eplusmunu}
&=&-\frac{D^{\mu}D_{\nu}d}{d}\\
&& -\frac{1}{2d L}\left[
\dd\left(
\tanh \frac{\dd}{2}-\tan \frac{\dd}{2}
 \right) \delta^\mu _\nu \right.\notag \\
&& \left.
 +\frac{1}{\dd}
 \left(
\tanh \frac{\dd}{2}+\tan \frac{\dd}{2}
 \right) F^\mu _\nu
\right]\notag\\
&&
+\frac{1}{d
  L}\mathcal{O}\hspace{-3pt}\left(\frac{d}{L}\right),\notag\\
\text{with}&& \hspace{-15pt}\dd \equiv \sqrt{- g^{(+)\,\mu
\nu}
\partial_{\mu}d \partial_{\nu}d}\notag\\
\text{and}&&\hspace{-15pt}  F^\mu _\nu=2 \partial^{\mu}d\,  \partial_{\nu}d -
\left(\partial d \right)^2 \delta^{\mu}_{\nu}, \notag
\ea
here the covariant derivative and the raising of indices are  performed
with respect to
 $g^{(+)}_{\mu \nu}=q_{\mu \nu}(y=0)$.
Note that, because of the modulus signs, we have
\ba
\left(\partial d\right)^2\equiv g^{(+)\mu\nu}\partial_\mu d
\partial_\nu d=-\dd^2.
\ea
We may note that these functions are well defined only for $\dd <
\pi$. 
 For $\dd \ge
\pi$, the inverse of the power expansion on the left hand side of
(\ref{expansion1}, \ref{expansion2}) is
ill defined. For the purpose of this study we shall therefore
restrict ourselves to the case where $\dd <\pi$ for which the expressions
(\ref{eplusmunu}) is well defined. Whether or not the theory
possesses actual velocity-dependent divergences as $\dd\rightarrow
\pi$ would be interesting to investigate with numerical
simulations.

All this analysis could be repeated using the negative-tension
brane as reference brane instead. This would be equivalent to
taking $-d$ instead $d$ all the way through.
 We would have then obtained the induced
Weyl tensor on the negative-tension brane $E^{(-)
  \mu}_{\phantom{(-)}\, \nu}$ as
\ba
\hspace{-10pt}E^{(-) \mu}_{\phantom{(-)}\, \nu}\hspace{-5pt}
&=&-\frac{D^{\mu}D_{\nu}d}{d}\\
&& +\frac{1}{2d L}\left[
\dd\left(
\tanh \frac{\dd}{2}-\tan \frac{\dd}{2}
 \right) \delta^\mu _\nu \right.\notag \\
&& \left.
 +\frac{1}{\dd}
 \left(
\tanh \frac{\dd}{2}+\tan \frac{\dd}{2}
 \right) F^\mu _\nu
\right]\notag\\
&&
+\frac{1}{d L}\mathcal{O}\hspace{-3pt}\left(\frac{d}{L}\right),\notag
\ea
where the covariant derivative and the raising of indices are
now with respect to $g^{(-)}_{\mu
\nu}=q_{\mu \nu}(y=1)$. This procedure has allowed us to
derive the Weyl tensor on both the positive- and negative-tension branes in a
covariant way. These expressions are exact in the close-brane limit.
 For simplicity we exposed in detail the case for
which $A^2=g_{yy}$ was assumed to be $y$-independent. It is unclear
whether such a gauge can always be fixed, but our final answer is
gauge-invariant and we show in Appendix \ref{A(y)} that taking $A$
to be $y$-dependent does not in fact affect the final result. For a
discussion of this gauge issue see \cite{Charmousis:1999rg}
%
%
%
\section{Theory in the close brane limit}
\label{4d theory small d}
\subsection{Modified Einstein equation
}
Using the result of the previous Section, in the close brane limit,
the positive-tension brane geometry will satisfy the following
modified Einstein equation:
\ba
G^{(+)}_{\mu \nu}&=&-E_{\mu \nu}^{(+)}\notag\\
&=&\frac{D_{\mu}D_{\nu}d}{d}\label{Einstein small d}
+\frac{1}{d L}\left[ \dd  \tanh \frac{\dd}{2}
 g^{(+)}_{\mu \nu} \right. \\
&& 
+\left.
\frac{1}{\dd}
 \left(
\tanh \frac{\dd}{2}+\tan \frac{\dd}{2}
 \right) \partial_{\mu}d \partial_{\nu}d\right]
\notag \\
&& +\frac 1 d \mathcal{O} (\frac d L).   \notag
\ea
describing gravity coupled to a scalar field in a
non-trivial way.
This theory has some higher-derivative corrections as expected but,
remarkably, they are both simple in form and entirely \emph{first
  order}, involving only powers of first derivatives. This leads to
the important result that the theory remains second order in
derivatives even when correct to \emph{all} orders in velocity.
The initial data on a Cauchy surface that needs to be specified for the
theory to be solved is the same as that needed
in the low-energy effective theory. There will be no need to
specify extra information or to consider the corrections to be
small, a feature which will make the theory straightforward to solve.

In order to solve this theory, we need to specify the equation of
motion of the scalar field. So far the traceless property of the Weyl
tensor has not been used, which gives rise to the modified Klein Gordon equation for the radion:
\ba
\Box d &=& \frac{\dd}{L}\left[
\tan\frac{\dd}{2}-
3\tanh \frac{\dd}{2}\label{KG small d}
 \right]\\
&&+\frac1 L \mathcal{O}(\frac d L). \notag
\ea
Using this equation of motion for the scalar field and the modified
Einstein equation (\ref{Einstein small d}),
the geometry on the brane can be found in the close-brane limit. We
may point out that only four-dimensional quantities are  involved.
Although this theory does not seem to be derivable from a
four-dimensional action, it may be solved using standard
four-dimensional methods.

\subsection{Low-energy limit}
In the low-energy limit, the four-dimensional effective theory presented
in Section \ref{section 4d eff} is exact. For consistency we may check that,
in the small velocity limit, we recover the small distance limit of
this theory.
To leading order in velocity, we may use
$
\tan \frac{\dd}{2}\simeq\tanh \frac{\dd}{2}\simeq\frac{1}{2}\dd
$
in (\ref{Einstein small d}) and (\ref{KG small d}), giving rise to the Einstein equation and
the equation of motion for the scalar field
\ba
G^{(+)}_{\mu \nu} \hspace{-5pt}&=&\hspace{-5pt}\frac{D_{\mu}D_{\nu}d}{d} +\frac{1}{d
L}\left(
\partial_{\mu}d \partial_{\nu}d
 -\frac 1 2 \left(\partial d \right)^2
 g^{(+)}_{\mu \nu} \right)\nn\\
\Box d&=&\frac{\left(\partial d\right)^2}{L},
\nn\ea
which is precisely the leading order in $d/L$ of the effective theory (\ref{Einstein eff in
d}, \ref{box d eff}). The equation of motion for the scalar field is
actually exact to all orders in $d/L$.

We might think that the low-energy effective theory (\ref{Einstein eff in d})
could still give a correct answer provided that the relation between $\Psi$
and $d$ was modified. If that was the case, the relation between these
two variables should include some higher order velocity terms: $\Psi=\Psi (d, \partial
d)$ such that, in the low-velocity limit, $\Psi (d, \partial d)\sim
e^{-d/L}$. The theory should then include some
terms of higher order in derivative of
the form $\partial^\alpha d D_\mu D_\nu \partial _\alpha
d\ \Psi_{, \partial d}$. This is not compatible with the theory in (\ref{Einstein small
d}) which only contains terms up to second order in derivatives. The
close limit theory in (\ref{Einstein small d}) and
the low-energy effective theory (\ref{Einstein eff in d}) are therefore
genuinely different.
%
%
\subsection{Conservation of Energy}
\label{section Conservation of Energy}
In order for the theory to be consistent, the Bianchi identity on
the brane needs to be preserved. This means that, to leading order in
$d/L$, the traceless condition (\ref{KG small d}) must imply the
right hand side of (\ref{Einstein small d}) to be transverse. In
other words, if we consider the right hand side of (\ref{Einstein small
d}) to be the stress-energy tensor of the scalar field $d$, the Klein
Gordon condition should impose that it is conserved, i.e. divergenceless. Using the equation of motion
for the scalar field, we may rewrite (\ref{Einstein small d}) and
(\ref{KG small d}) in the form:

\ba
G_{\mu \nu}^{(+)}&=&
T_{\mu \nu} \notag \\
T_{\mu \nu}&=&\frac{D_{\mu}D_{\nu}d}{d}+\frac{1}{L d}
\left(f(z) g_{\mu \nu}^{(+)}+g(z) F_{\mu \nu}\right)\label{Tuv=}\\&&-\bar
E_{\mu \nu} \nn \\
&=&\frac{1}{d}\Big[
D_{\mu}D_{\nu}d
+\frac{2}{L}g\, \partial_{\mu}d\partial_{\nu}d
-d\, \bar E_{\mu \nu} \label{Tuv2} \\
&&-\left(
\Box\, d +\frac{z^2 g}{L}+\frac{3f}{L}-d\, \bar E
\right)g_{\mu
\nu}^{(+)} \notag
\Big]\\
\Box \, d&=&-\frac{1}{L}\left(4f+2z^2 g\right)+ d\,  \bar E,
\label{box d 2}
\ea
where for simplicity we used the notation
$z=\dd$, $f\equiv f(z^4)=\frac{z}{2}\left(\tanh \frac{z}{2}-\tan \frac{z}{2}
 \right)$ and $g\equiv g(z^4)=\frac{1}{2z}\left(\tanh \frac{z}{2}+\tan \frac{z}{2}
 \right)$. $f$ and $g$ are viewed as functions of $z^4=\left(\partial
 d\right)^4$ since their Taylor expansions only contain powers of
 $z^4$. Therefore, for example,
\ba
\partial_\mu f(z)=2z^2f'(z)\, \partial_\mu \left(z^2\right).
\ea
 We have as well added the next-order correction $d \bar E_{\mu \nu}$ where $\bar E_{\mu \nu}\sim
 d^0$. As we shall see, although $\bar E_{\mu \nu}$ does not
 contribute to the leading order of the Einstein equation, it does
 contribute to the leading order of the divergence of $T_{\mu \nu}$.
 This is due to the fact that, although $d \bar E _{\mu \nu}\sim d$
 is negligible, its derivative is not $\left(\partial_\mu d  \right) \bar E^\mu _{\nu}\sim
 d^0$. We therefore need to consider its contribution as well. The
 transverse requirement will therefore impose a condition on the
 next to leading order in $d$.
Using the result of Appendix \ref{appendix div T},
 the divergence of the stress-energy tensor as defined in (\ref{Tuv=}) is given
 by:
 \ba
 d D_\mu T^\mu_\nu&=&
 -\frac{12}{L^2}\left[g f+z^2\left(f-z^2 g\right)\left(f'+z^2
 g'\right)\right]\partial_\nu d\notag \\
 &&- \left[\bar E^\mu _{\nu}-\bar E \delta^\mu_\nu
 \right]\partial_\mu d+\frac{1}{L^2}\mathcal{O}(d/L)
 \label{Du Tuv} \\
 &=&0. \notag
 \ea
 For the stress-energy to be transverse to leading order in $d$ we
 therefore have a constraint equation on the next-to-leading order
 contribution $\bar E_{\mu \nu}$. Furthermore, from the low-energy
 effective theory (using (\ref{Einstein eff in d})), to leading
 order in velocities and in $d/L$, $\bar E_{\mu \nu}$ should be:
 \ba
\bar E_{\mu \nu}\hspace{-5pt}&=&\hspace{-5pt}\frac{1}{L}\left[
D_\mu D_\nu d
+\frac{1}{L}\left(\partial_{\mu} d \partial_{\nu} d-\frac{1}{2}
\left(\partial d\right)^2g_{\mu\nu}^{(+)} \right)
\right] \label{bar E eff} \\
&&\hspace{-5pt}+\frac{1}{L^2}\mathcal{O}\left(\left(\partial
d\right)^4, d/L
\right). \notag
 \ea
A natural ansatz for $\bar E_{\mu \nu}$ is therefore:
\ba
\bar E_{\mu \nu}=\frac{1}{L}\left[
D_\mu D_\nu d
+\frac{1}{L}\left(\bar f(z^4) g_{\mu\nu}^{(+)}+\bar g (z^4) F_{\mu \nu} \right)
\right]. \label{anstaz E bar}
\ea
Using this ansatz in the constraint (\ref{Du Tuv}), we find the
equations for $\bar f$ and $\bar g$:
\ba
\bar f -f+z^2(\bar g -g)&=&4\left(gf +z^4f g'-z^4 g f'\right)\\
&+&4\, z^2 \left(f f'-z^4g g' \right). \nn
\ea
Since $\bar f$ and $\bar g$ are functions of $z^4$ uniquely, their
 expression is:
\ba
&& \bar f(z^4)=f+4g f+4z^4\left(f g'-g f'\right)\label{bar f}\\
&& \bar g(z^4) = g +4 f f'-4z^4 g g'. \label{bar g}
\ea
In the slow-velocity limit, we may check that $\bar f =\mathcal{O}(\left(\partial
d\right)^4)$ and $\bar g=\frac{1}{2}+\mathcal{O}(\left(\partial
d\right)^4)$ so we recover the result from the low-energy effective
theory (\ref{bar E eff}). \\ Furthermore, using the expression (\ref{Tuv=}),
the next to leading order
$\bar E_{\mu \nu}$ (\ref{anstaz E bar}) will be conformally related
to the leading order $\bar E_{\mu \nu} =\frac d L E_{\mu \nu}^0$
only if $\bar f=f$ and $\bar g=g$.
But from the expressions (\ref{bar f}, \ref{bar g}),
this can only be the case in the low-velocity limit
where $\bar f = f = -\frac{z^4}{4!}+\mathcal{O}  (z^8)$ and
$\bar g =g = \frac{1}{2}+\mathcal{O}  (z^4)$.
The higher order terms in velocities do not match and the
conformal relation $\bar E_{\mu \nu} =\frac d L E_{\mu \nu}^0$ does
not hold.

This expression for $\bar E_{\mu \nu}$ with the
relations (\ref{bar f}, \ref{bar g}) is consistent with the
close limit theory since it ensures that the stress-energy tensor is
conserved to leading order in $d/L$.
However we may emphasise that this is not the only possible answer and
relies strongly on the ansatz (\ref{anstaz E bar}). Since we
only want to focus in the leading order in $d/L$ of the theory, the
exact expression of $\bar E_{\mu \nu}$ will not be relevant, it
 is enough to show that it is possible to introduce its contribution in such
a way that the transverse condition is preserved. If we wanted to go
further in the study of the next to leading order term of the
theory, we should check that this expression of $\bar E_{\mu \nu}$
is consistent with its exact expression that can be derived in the
background using the SAdS five-dimensional geometry. This is beyond the purpose of
this work which aims to focus on the close brane limit and to
neglect any terms which are not of leading order in the expansion.

\subsection{Consistency check for the background}
Having a theory which should be exact
in the close brane limit, we may now check that it correctly
reproduces the exact background behaviour obtained in (\ref{a' exact}, \ref{H
exact}). Taking the positive-tension brane to be endowed with a flat
FRW induced metric with scale factor
$a_+(t)$, the Einstein equations (\ref{Einstein small d}) reduce to
\ba
\label{cc1}
3H_+^2&=&\frac{\ddot{d}}{d}+\frac{\dot{d}}{Ld}\tan{\frac{\dot{d}}{2}}\\
\label{cc2}
2\dot{H_+}+3H_+^2&=&H_+\frac{\dot{d}}{d}-\frac{\dot{d}}{Ld}\tanh{\frac{\dot{d}}{2}},
\ea
and the Klein Gordan equation (\ref{KG small d}) gives
\be
\label{cc3}
\ddot{d}+3H_+\dot{d}=\frac{\dot{d}}{L}\left(3\tanh\frac{\dot{d}}{2}-\tan\frac{\dot{d}}{2}\right).
\ee
Combining (\ref{cc1}) and (\ref{cc3}) gives
\be
\label{cc4}
H_+^2=\frac{\dot{d}}{d}\left(\frac{1}{L}\tanh\frac{\dot{d}}{2}-H_+\right).
\ee
Also we have immediately that $a_+(t)$ is linear in conformal time
(which is merely a consequence of the induced metric being FRW with
empty branes). As with the analogous low-energy result
(\ref{quadratic}), (\ref{cc4}) has two solutions, one of which
has $\dot d=0$ at the collision and for which both $\dot d$ and $H_+$
are finite and non-vanishing. Again, it is the latter we
consider since this is the one corresponding to branes moving in
opposite directions. Therefore for this study both $H_+$ and $a_+$ are finite in the limit $d\rightarrow 0$,
giving the leading order behaviour of $H_+$ from (\ref{cc4}) as
\be
\label{H+ allv}
H_+=\frac{1}{L}\tanh\frac{\dot{d}}{2}+\mathcal{O}  \left(\frac{d}{L^2}\right).
\ee
Replacing $d$ with $-d$ will give the corresponding result for the
negative-tension brane:
\be
\label{H- allv}
H_-=-\frac{1}{L}\tanh\frac{\dot{d}}{2}+\mathcal{O}  \left(\frac{d}{L^2}\right).
\ee
The modified Friedmann equations on both branes
(\ref{H+ allv}, \ref{H- allv}) are precisely the exact result (\ref{H
  exact}) we obtained from solving the
five-dimensional equations in Section \ref{5d}.
This is a non-trivial check on the validity of this close-brane four-dimensional
theory as it reproduces the right result to all order
in velocities in the small distance limit and not only to leading
order in $\dot d$ as was the case for the low-energy effective theory.

%
%
\section{Extensions}
\label{extension}
\subsection{Introduction of a potential to low-energy effective theory}
\label{add potential}
%
%
The low-energy presented in Section \ref{section 4d eff} was expressed
in the positive-tension brane frame. In order to add a potential
to this theory, it is more natural to  work in the Einstein frame.
The Einstein frame is related to the brane frame by the conformal transformation:
\begin{eqnarray}
g^{(+)} _{\mu \nu}&=&\left(\cosh \left(\phi / \sqrt{6} \right) \right)^2
\bar{g}_{\mu \nu},
\label{g+intermasofg4d} \\
g^{(-)} _{\mu \nu}&=&\left(-\sinh \left(\phi / \sqrt{6} \right) \right)^2
\bar{g}_{\mu \nu}.
\end{eqnarray}
where any ``bar" quantity designates a quantity with
respect to the Einstein frame.
The conformal scale factor in (\ref{Einstein eff in d}) is related to
the scalar field
$\phi <0 $ by
$\Psi=e^{-d/L}=-\tanh \left(\phi / \sqrt{6} \right)$.
In that frame, the four-dimensional effective theory can be
derived from the effective action of a scalar field minimally
coupled to gravity:
\begin{eqnarray}
S=\int d^4 x\sqrt{-\bar g} \left( \frac {1}{2} \bar R-\frac{1}{2}
\left(\partial \phi \right)^2 \right),
\label{4daction}
\end{eqnarray}
giving rise to the equation of motion for the scalar field:
$\bar{\Box} \, \phi= 0$, Cf.\cite{Kanno:2002ia,Barvinsky:2001tm}.

It is possible to modify this effective theory by adding a potential by
hand in (\ref{4daction}):
\begin{eqnarray}
S=\int d^4 x\sqrt{-\bar g} \left( \frac {1}{2} \bar R-\frac{1}{2}
\left(\partial \phi \right)^2 -U(\phi)\right).
\label{4daction+V}
\end{eqnarray}
In that case the equation of motion for the scalar field is modified
to:
\ba
\bar{\Box} \, \phi= U'(\phi),
\ea
and the Einstein equation in Einstein frame becomes:
\ba
\bar G_{\mu \nu}=\partial_\mu \phi \partial_\nu \phi
-\left(\frac{1}{2}(\partial \phi)^2+U \right)\bar{g}_{\mu \nu}.
\ea

We can now go back to the positive-tension brane frame by performing the
conformal transformation (\ref{g+intermasofg4d}) with
$\left(\cosh \left(\phi / \sqrt{6} \right) \right)^2 \simeq
\frac{L}{2d}$.  To linear order in $d/L$, the
theory on the brane is modified to:
\ba
G^{(+)}_{\mu \nu}&=&
\frac{1}{d}\Big[
D_{\mu}D_{\nu}d-\Box d\,  g_{\mu \nu}^{(+)} \label{Guv eff V}\\
&& +\frac{1}{L}\left(\partial_\mu d\, \partial_\nu d
+ \frac{1}{2}\left(\partial d \right)^2 g_{\mu \nu}^{(+)}\right)
-V(d) \,  g_{\mu \nu}^{(+)}
\Big]\notag \\
\Box \, d &=&\frac{\left(\partial d \right)^2}{L}
+\frac{2}{3}\left(d\, V'(d)-2 V(d) \right), \label{box d eff V}
\ea
where all covariant derivatives and index raising is taken with
respect to $g_{\mu \nu}^{(+)}$ and for simplicity we used the
potential $V(d)$ defined as:
\ba
V(d)=d \left(1-\Psi^2 \right)U \simeq \frac{d^2}{L}\, U.
\ea
The modification of the equation of motion of the scalar field
(\ref{box d eff V}) ensures the right hand side of the Einstein
equation (\ref{Guv eff V}) to stay conserved even after the addition
of a potential.
The same procedure can now be applied to the theory in the close brane limit.
%
%
\subsection{Addition of a potential in the close brane theory}
\label{section small d with V}
We consider the addition of a potential in the close brane theory by
modifying the stress-energy tensor in the Einstein equation (\ref{Tuv2}):
\ba
T_{\mu \nu}
&=&
\frac{1}{d}\Big[
D_{\mu}D_{\nu}d
+\frac{2}{L}g\, \partial_{\mu}d\partial_{\nu}d
-d\, \bar E_{\mu \nu} \label{Tuv + V} \\
&&-\left(
\Box\, d +\frac{z^2 g}{L}+\frac{3f}{L}-d\, \bar E +V(d)
\right)g_{\mu
\nu}^{(+)} \notag
\Big].
\ea
For this stress-energy to remain conserved after the addition of the
potential, the equation of
motion of the scalar field (\ref{box d 2}) has to be modified as
well:
\ba
\Box \, d&=&-\frac{1}{L}\left(2z^2 g+4f\right)+ d\,  \bar E +W,
\label{box d +V}
\ea
where the correction term $W$ is a functional of the potential
$V(d)$ to be determined. The modified Klein Gordon equation (\ref{box
  d +V}) should be consistent with the conservation of the modified
stress-energy tensor in (\ref{Tuv + V}).
The divergence of $T_{\mu \nu}$ is calculated in the appendix
\ref{appendix div T}
using the constraint (\ref{bar E eff}) for $\bar E_{\mu \nu}$:
\ba
d\,  D_\mu T^\mu_\nu&=&
\partial_\nu d \Big[
-V'(d)+\frac{3W+4V}{2d}
+\frac{2g}{L}\left(3W+2V \right) \notag\\
&& \hspace{-25pt}+\frac{12}{L}z^2\left(f'+z^2 g'\right)\left(V+W
\right)
\Big]+\frac{1}{L^2}\mathcal{O}  (d/L).
\ea
To lowest order in $d/L$ we therefore need to modify the equation for
the $d$ with a term $W$:
\ba
W=\frac{2}{3}\left(d V'(d)-2 V(d) \right).
\ea
This is precisely the same term which had to be introduced in the
equation (\ref{box d eff V}) for $d$ in the low-energy theory. This
procedure is therefore completely consistent with the low-energy theory
and will give the same result to leading order in $\left(\partial d
\right)$.
It is possible to modify the close brane theory by introducing a
potential, in such a way that the conservation of energy and thus the
Bianchi identity remain unaffected.

%
%

\subsection{Detuned tensions}
A full analysis of how the argument of Section \ref{4d theory small d} would run in the presence of
matter on the branes is beyond the scope of this paper. We can,
however, consider the simple example of brane cosmological constants
where the stress-energy tensors are proportional to the induced metric
to get an idea for how the coupling to matter might
look. Specifically, we allow now for a detuning of the tensions to
\ba
 \lambda_+ = \frac{6}{L}+\sigma_+\qquad  \lambda_- =
-\frac{6}{L}+\sigma_-.
\ea
The analysis is only slightly modified from the above and
we shall only sketch it here for brevity. Using (\ref{proj einstein
  gen}), (\ref{taylor2})
is modified to
\ba
\sum_{n\ge 1} \frac{1}{n!}
\  ^{0}\! K^{\mu\, (n)}_{\, \nu}=\frac{1}{6}\left(\sigma_++\sigma_-\right)\delta^\mu_\nu,
\ea
which gives rise to a more complicated version of (\ref{operators}):
\ba
K^{\mu
  '}_\nu\left(y=0\right)&=&\left(\frac{1}{L}+\frac{\sigma_+}{6}\right)
\left[\sqrt{\hat{O}}\tanh\left(\frac{\sqrt{\hat{O}}}{2}\right)\right]\delta^\mu_\nu
\nn\\
&&+\frac{\sigma_++\sigma_-}{6}\frac{\sqrt{\hat{O}}}{\sinh{\sqrt{\hat{O}}}}\delta^\mu_\nu.
\ea
The analogous result to (\ref{eplusmunu}) is then
\ba
\hspace{-10pt}E^{(+) \mu}_{\phantom{(+)}\, \nu}\hspace{-5pt}
&=&-\frac{1}{d}K^{\mu\, \prime}_{\,
\nu}(y=0)-\frac{D^{\mu}D_{\nu}d}{d}
-\frac{\delta^\mu_\nu}{3}\left(\frac{\sigma_+}{L}+\frac{\sigma_+^2}{12}\right),\notag\\
 &=&-\frac{D^{\mu}D_{\nu}d}{d}\\
&& -\frac{1}{2d}\left(\frac{1}{L}+\frac{\sigma_+}{6}\right)\left[
\dd\left(
\tanh \frac{\dd}{2}-\tan \frac{\dd}{2}
 \right) \delta^\mu _\nu \right.\notag \\
&& \left.
 +\frac{1}{\dd}
 \left(
\tanh \frac{\dd}{2}+\tan \frac{\dd}{2}
 \right) F^\mu _\nu
\right]\notag\\
&&
+\frac{\left(\sigma_++\sigma_-\right)}{3}\Big[
\dd\Big(
\cosech \dd+\cosec \dd\Big) \delta^\mu _\nu \notag \\
&& \left.
 +\frac{1}{\dd}
 \Big(
\cosech \dd-\cosec \dd
 \Big) F^\mu _\nu
\right]\nn\\
&&
-\frac{\delta^\mu_\nu}{3}\left(\frac{\sigma_+}{L}+\frac{\sigma_+^2}{12}\right)
+\frac{1}{d
  L}\mathcal{O}\hspace{-3pt}\left(\frac{d}{L}\right).\notag
\ea
The Einstein and Klein-Gordon equations are then modified to include
complex couplings of the tensions to both the radion $d(x)$ and powers of
its first (but not higher) derivative.
\ba
G^{(+)}_{\mu \nu}&=&-E^{(+)}_{\mu\nu}-g^{(\pm)}_{\mu\nu}\left(\frac{1}{L}\sigma_++\frac{1}{12}\sigma_+^2\right)\notag\\
&=&\frac{D_{\mu}D_{\nu}d}{d}
+\frac{1}{d}\left(\frac{1}{L}+\frac{\sigma_+}{6}\right)\left[ \dd \tanh \frac{\dd}{2}
 g^{(+)}_{\mu \nu} \right.\nn \\
&& 
+\left.
\frac{1}{\dd}
 \left(
\tanh \frac{\dd}{2}+\tan \frac{\dd}{2}
 \right) \partial_{\mu}d \partial_{\nu}d\right]
\notag \\
&&+\Sigma_{\mu\nu} +\frac 1 d \mathcal{O} (\frac d L)\\
\Sigma_{\mu\nu}&=&\frac{\left(\sigma_++\sigma_-\right)}{6d}\Big[\dd\cosech\dd
  g^{(+)}_{\mu\nu}\nn\\
&&+\frac{1}{\dd}\Big(\cosech\dd-\cosec\dd\Big)\partial_\mu d
  \partial_\nu d\Big]\nn\\
&&-\frac{2}{3}g^{(+)}_{\mu\nu}\left(\frac{\sigma_+}{L}+\frac{\sigma_+^2}{12}\right)
\ea

\ba
\Box d &=& -\left(\frac{1}{L}+\frac{\sigma_+}{6}\right)\dd\left[
3\tanh\frac{\dd}{2}-
\tan \frac{\dd}{2}\right]\nn\\
&&
-\frac{\left(\sigma_++\sigma_-\right)}{6}\dd\Big[3\cosech\dd+\cosec\dd\Big]\nn\\
&&-\frac{4d}{3}\left(\frac{\sigma_+}{L}+\frac{\sigma_+^2}{12}\right)+\frac1
L \mathcal{O}(\frac d L).
\ea
In the low-energy limit, these reduce to
\ba
G^{(+)}_{\mu \nu} \hspace{-5pt}&=&\hspace{-5pt}\frac{D_{\mu}D_{\nu}d}{d} +\frac{1}{d
L}\left(
\partial_{\mu}d \partial_{\nu}d
 -\frac 1 2 \left(\partial d \right)^2
 g^{(+)}_{\mu \nu} \right)\\
&&
+\frac{\left(\sigma_++\sigma_-\right)}{6d}g^{(+)}_{\mu\nu}+\frac{1}{d}\mathcal{O} \left(d/L,\partial
d^4,\sigma_\pm\partial d^2\right)\nn\\
\Box d&=&\frac{\left(\partial d\right)^2}{L}-\frac{2}{3}\left(\sigma_++\sigma_-\right).
\ea
which agree with the standard results of the low-energy effective
theory in the presence of detuned tensions \cite{Webster:2004sp}.

\section{Conclusion}
\label{conclusion}
In the first part of this work we pointed out the discrepancy between
predictions from the effective four-dimensional low-energy theory and
the exact five-dimensional results.  The difference has been
established in the regime where the branes were close and all through
this paper we worked to leading order in $d/L$.

In order to go beyond the low-energy effective theory, we established a
formalism to find the electric part $E_{\mu \nu}$
of the five-dimensional Weyl tensor on both branes. This tensor
represents the only quantity which is unknown from a four-dimensional
point of view as it encodes information about the bulk
geometry. Finding its expression on the brane is therefore the key
element in order to study the geometry of a brane within an extra
dimension.

Using this formalism in the small distance limit, we found an exact
expression for the Weyl tensor on each brane, valid at leading order
in $d/L$ but at all orders in velocities, or for any energy scale.
We were therefore able to modify the regime of validity of the effective
theory from a low-energy regime valid independently of the branes
distance to a regime valid at all energies for close branes.
This regime of validity is relevant for cosmology since
it represents one of the main focus of present braneworld
models. Understanding the behaviour of branes just before or just
after a collision is indeed crucial if our Universe emerged from a
brane collision. If the Big Bang were initiated by such a
collision, it is consistent to assume that the large scale
structure was produced in a regime where both branes were close.
Even if this regime is not valid after a while, effects
that originated during this period are unlikely to be eliminated
once the branes are far apart and the fifth dimension has opened up,
leaving the possibility of exciting the Kaluza Klein modes.
We therefore believe that understanding the consequences of
cosmologies in this limit will allow us to study the viability of such scenarios by
comparing them with observational data
\cite{Calcagni:2003sg,Seahra:2005us,Liddle:2003gw,Ashcroft:2002ap,
Calcagni:2004bb,Tsujikawa:2003zd}.

In this paper we have established a theory that allows us to study
such scenarios and have checked the consistency of its predictions for
homogeneous and isotropic geometries.
We argue that this theory will be remarkably straight-forward to use
since it includes only four-dimensional quantities and is effectively
second order in derivatives, the only higher-order corrections coming
as powers of first derivatives. This feature will facilitate any
comparison with other four-dimensional theories.
 It is different from other
higher-derivative theories
\cite{deRham:2004yt,Mukohyama:2001jv,Kanno:2003vf} in that it is not
derived using the assumption that it can be derived from an action,
and it is different from a theory relying on purely string effects
\cite{deBerredo-Peixoto:2004if,Kanno:2004hr,Silverstein:2003hf}.
Already for the background, an
interesting result can be pointed out: when the branes are empty the Hubble parameter on
each brane is bounded by $L^2 H^2 <1$, which could not have been derived from the low-energy effective
theory.
This does not hold anymore when
some matter is introduced on the branes or if a potential is added for
the radion.
Another interesting result of the theory is that the expansion seems to
break down when the velocities are of order $\left|\partial d \right| \sim \pi$. This is a
feature which has never been pointed out before and we believe it
suggests the presence of interesting physics which might be
worth studying.

To extend the theory, we considered the addition of a
potential for the radion. This will be interesting if we want to study curvature
perturbations as the origin of the large scale structure. In order to
begin to understand the way matter on the brane would be coupled in this
theory we have extended the model to the case where the brane
tensions are detuned. Both these extensions will be useful for further progress
in the understanding of braneworld cosmology after or before a collision.

\section*{Acknowledgements}
We would like to thank Anne Davis for supervising this work and for
her comments on the manuscript. SLW is
supported by PPARC and would like to thank James O'Dwyer and David
Richards for pointing out the breakdown of the expansion as
$\dd\rightarrow\pi$. CdR is supported by DAMTP and would like to thank
Andrew Tolley for useful discussions throughout the development of
this paper.


%
%
\appendix
%
%
\section{Leading order Derivative of the extrinsic Curvature}
\label{appendix n case}
\subsection{Gauge where A is y-independent}
The purpose of this Section is to formulate the recursive expression
for the derivatives of the extrinsic curvature on the brane in the
regime where the branes are close $d\ll L$.

We recall from Section {\ref{n=1 and n=2}} that on the positive
tension brane, for $y=0$, we have $K^{(m)} \sim d^{0}$ for
$m=0,1,2$. Furthermore, from (\ref{gamma '}), at $y=0$, we have
$\Gamma' \sim \partial d\sim d^{0}$. We may as well
calculate
its second derivative:
\ba
\Gamma ^{\, \alpha\ \prime \prime}_{\mu \nu}(y)&=&
(d\, K^{\alpha\, \prime}_{\, \nu})_{;\mu}
+(d\, K^{\alpha\, \prime}_{\, \mu})_{;\nu} -(d \, K^{\prime}_{\mu
\nu})^{;\alpha} \label{gamma ''} \\
&& +d\,\left( \Gamma^{\, \alpha \, \prime}_{\mu \rho}
K^\rho_\nu+\Gamma^{\, \alpha \, \prime}_{\nu
\rho}
K^\rho_\mu
-2 \Gamma^{\,\rho \, \prime}_{\mu \nu} K^\alpha_\rho \right)\nn \\
&& -d\, q^{\alpha \beta}q_{\mu \sigma}\,  \left(
\Gamma^{\, \sigma \, \prime}_{\beta \rho} K^\rho_\nu
-\Gamma^{\, \rho \, \prime}_{\beta \nu} K^\sigma_\rho
\right) \notag\\
&&-d\,\left(q^{\alpha \beta}q_{\mu \sigma}\right)^{\prime}\,
\left(
\Gamma^{\, \sigma}_{\beta \rho} K^\rho_\nu
-\Gamma^{\, \rho }_{\beta \nu} K^\sigma_\rho
\right).\nn
\ea
On the brane the last line is of order $(d/L)^2$ and the two
middle ones are of order
$d/L$. They are therefore negligible.
The first line gives a contribution of order $\Gamma ^{\prime \prime}(0)
\sim \partial d K'(0)
\sim d^{0}$:
\ba
\Gamma ^{\, \alpha\, \prime \prime}_{\mu \nu}(0)=
\partial_\mu d\, K^{\alpha\, \prime}_{\, \nu}(0)
+\partial_\nu d\, K^{\alpha\, \prime}_{\, \mu}(0)
-\partial^\alpha d \, K^{\prime}_{\mu \nu}(0),
\ea
where terms of order $\frac{d}{L}$ have been omitted.

In what follows we will use a symbolic notation omitting any indices
or coefficients. We write symbolically, $q_{\alpha
\beta}=q$, $K^{\alpha}_{\beta}=K$, $\partial_{\alpha}=\partial$ and
$\partial_y=\, '$. In particular, we have: $q'=dqK$.
Using this notation and expressing $E^\mu _{\nu}$ in terms of $K^{\prime
\mu}_{\
\nu}$ in (\ref{E prime}), we may symbolically write an equation for
$K^{\prime \prime \mu}_{\ \ \nu}$:
\ba
K^{\prime \prime}&=&
d\, \left(\partial^2 +\Gamma \partial +\partial \Gamma +\Gamma^2 \right)
 q'
+d K^3 +d K \label{K'' symbolic}\\
&& +d K K'+q\,  \partial d \left(\Gamma' +d\, \Gamma K +d\, \partial K
\right)+ \partial^2 d\, q', \notag
\ea
where again by $\partial$ we designate a normal derivative (as opposed to covariant
derivative) with respect to any coordinate $x^{\mu}$ along the four-dimensional
$y=\text{const}$ hypersurface. This expression is true for any $y$.
The leading order in $d/L$ in this expression comes from the term $q\,
\partial d\, \Gamma' \sim d^0$. All the other terms are of order $d$.

For $n=2$, we therefore have on the brane:
\ba
K^{(m)}(0) \sim \Gamma^{(m)}(0) \sim d^{0} \ \ \ \forall \ m \le n .
\ea
Now let us assume that this relation holds for a given $n=l+1$. In
particular,
$\Gamma^{(l+1)}(0) \sim d^{0}$ and
\ba
&& q^{(m+1)}(0)\sim d K^{(m)} d  \ \ \forall\  0\le m \le l+1,\\
&&  q (0)\sim d^{0}.
\ea
For the next order in $n+1$, we have:
\ba
&& K^{(n+1)}=\partial_y^{(l)}K^{\prime \prime} \label{K(l+2) symbolic}\\
&& \phantom{K^{(n+1)}}=
\partial_y^{(l)}
\Big[
d \left(\partial^2 +\Gamma \partial +\partial \Gamma +\Gamma^2 \right)
 q'
  \\
&&\phantom{K^{(n+1)}\partial_y^{(l)}}
+d K^3+d K +d K K'+ \partial^2 d\, q' \notag \\
&&\phantom{K^{(n+1)}\partial_y^{(l)}}+q\,  \partial d \left(\Gamma' +d\, \Gamma K +d\, \partial K
\right)
\Big], \notag
\ea
with
\ba
\left. \partial_y^{(l)} \, d\, \partial ^2 q'\right|_{y=0}= d \partial
^2\left[
q^{(l+1)}(0)\right] \sim d.
\ea
Similarly we may check that
\ba
\hspace{-25pt}\left.
\partial_y^{(l)}\left[d\, \left(\Gamma
\partial +\partial \Gamma +\Gamma^2 \right)
 q'+d\, K^3 +d\, K \right]\right|_{y=0}\hspace{-8pt}& \sim&\hspace{-2pt} d\hspace{3pt}\\
\hspace{-25pt} \left.\partial_y^{(l)}\left[d\, K K'+q
\partial d\, \left(d \Gamma K
+d \partial
K
\right)+ \partial^2 d \, q' \right]\right|_{y=0}\hspace{-8pt}& \sim &\hspace{-2pt} d,\hspace{3pt}
\ea
and
\ba
\left. \partial_y^{(l)}\left[q \Gamma' \partial d
\right]\right|_{y=0}= \partial d\, \sum_{m = 0}^{l}C^m_l
q^{(l-m)}(0)
\Gamma^{(l+1)}(0),\hspace{10pt}
\ea
with  $q^{(l-m)}(0) \Gamma^{(m+1)}(0) \sim d^{1} \text{ if } m < l $
and $q \Gamma^{(l+1)}(0) \sim d^{0}$ so that:
\ba
\left.\partial_y^{(l)}\left[q \Gamma' \partial d
\right] \right|_{y=0}\sim  d^{0}.
\ea
We have therefore shown that $K^{(n+1)}(0)\sim d^{0}$
 and its leading contribution comes from the derivative of the
Christoffel symbol uniquely.

We want now to show that the same is true for the Christoffel
symbol. We have $K^{(m)}(0) \sim d^{0}$ for any $m \le n+1$
and $\Gamma^{(m)}(0) \sim d^{0}$ for any $m\le n$.
Using the relation (\ref{gamma '}) for $\Gamma ^{\, \alpha\, \prime}_{\mu \nu}$, we have:
\ba
\Gamma ^{\, \alpha\, (n+1) }_{\mu \nu}(y)
=&&\hspace{-10pt}
 (d \, K^{\alpha\, (n) }_{\, \nu})_{,\mu}
+(d \, K^{\alpha \, (n) }_{\, \mu})_{,\nu} \notag -q_{\mu \rho}(d \,
 K^{\rho\, (n)}_{\, \nu})^{, \alpha} \\
&& \hspace{-20pt}
+d\, \partial_y^{(n)}
\Big[\Gamma^\alpha_{\mu \rho}
K^\rho_\nu+\Gamma^\alpha_{\nu \rho}
K^\rho_\mu
-2 \Gamma^\rho_{\mu \nu} K^\alpha_\rho
\Big.  \\
&& \hspace{-20pt} \phantom{+\frac{1}{d^n}\partial_y^{(n)}}
 \Big. -q^{\alpha
\beta}q_{\mu \sigma}\left(
\Gamma^\sigma_{\beta \rho} K^\rho_\nu
-\Gamma^\rho_{\beta \nu} K^\sigma_\rho
\right)\Big] \notag\\
&&\hspace{-20pt} -\,\sum_{m=0}^{n-1}
C^m_{n-1}
\left(d K^{\rho\, (m)}_{\, \nu}\right)_{,\,
\beta} \partial_y^{(n-m)}\left(q^{\alpha
\beta}q_{\mu \rho}\right),\notag
\ea
where
$
d\, \partial_y^{(n)}\left(\Gamma
K \right) \sim d$.
$
\partial_y^{(n-m)}\left(q^{\alpha
\beta}q_{\mu \rho}\right) \sim d$ for $ n > m$ so the last sum goes as
$d$.
Finally the
first term goes as: $(d \,
K^{\alpha\, (n)}_{\, \nu})_{,\mu}=
d_{, \mu} K^{
\alpha\, (n)}_{\, \nu}+\mathcal{O}(d)$.
The Christoffel symbol therefore goes as $\Gamma ^{(n+1)}\sim d^{0}$ for
all $n$ and its leading contribution is:
\ba
\Gamma ^{\, \alpha\, (n+1)}_{\mu \nu}(0)=
d_{,\mu} K^{ \alpha \, (n)}_{\, \nu}
+d_{,\nu} K^{\alpha\, (n)}_{\, \mu}
-d^{,\alpha} K^{(n)}_{\mu \nu}
\ea
where terms of order $\mathcal{O}  (d^1)$ have been omitted.
We can therefore finally conclude that
\ba
K^{(n)}\sim \Gamma^{(n)}\sim d^{0}\
\text{ for any } n \ge 0,
\ea
and the leading contribution for the extrinsic
curvature comes from the derivative of the Christoffel symbol only.
Indeed for $n=2$, we have:
\ba
K^{ \mu\, \prime \prime}_{\, \nu}(0)=q^{\beta \nu}
\Gamma^{\, \alpha\, \prime}_{\beta \nu}\  \partial_\alpha
d+\mathcal{O}\left(\frac{d}{L}\right),
\ea
and similarly for any $n\ge 2$,
the leading term in the extrinsic curvature comes
uniquely from the derivative of the Christoffel symbol:
\ba
 \left. K^{\mu \, (n)}_{\, \nu}\right|_{y=0}&=&q^{\beta \nu}
\Gamma^{\, \alpha\, (n-1)}_{ \beta \nu} \, \partial_\alpha
d+\mathcal{O}\left(\frac{d}{L}\right) \\
&=&
\left(
d^{,\mu} K^{\alpha\, (n-2)}_{\, \nu}
+d_{,\nu} K^{\alpha \mu\, (n-2)}\right.\\&& -\left.
d^{,\alpha} K^{\mu\, (n-2)}_{\, \nu}
\right)\, \partial_\alpha d
+\mathcal{O}\left(\frac{d}{L}\right).
\notag
\ea
We therefore have a recursive expression for the leading order in
$d/L$ of the normal derivatives of the extrinsic curvature that we
can use in order to formally sum the Taylor expansion (\ref{taylor1}) or
(\ref{taylor2}).

\subsection{Gauge where A depends on y}
\label{A(y)}
The previous result has been derived assuming that $A$ in (\ref{ds^2=})
was independent of $y$. It is possible to show that such a dependence
does not affect the final result.
In order to show this we will dissociate the derivatives of $A$ from
the other derivatives.
Then by summing over the derivatives of $A$, we will recover the
 quantity
$d$.
We denote by $\delta_A = A'(y)\partial_A + A''(y)\partial_{A'}
+ \partial_\alpha A'(y)\partial_{\partial_\alpha A}+\cdots$
the $y$-derivative which acts exclusively on $A$.

In order to differentiate any other quantity, we can apply the
operator $\bar{\partial}_y=\partial_y-\delta_A$ which has no effect on
$A$. In particular we write:
\ba
\bar{\partial}_y Q(y) =\partial_y Q(y) \equiv Q'(y),
\ea
for any quantity $Q=K^{\mu}_{\, \nu}, E^{\mu}_{\, \nu},q_{\mu \nu},
\Gamma^{\, \alpha}_{\mu \nu}$. In the previous Section, $A$ was
$y$-independent and so the operator $\delta_A$ had no effect, we
 simply had $\bar{\partial}_y=\partial_y$. Now this does not hold
anymore and we need to include the effect of $\delta_A$ in the Taylor
expansion (\ref{taylor1}) baring in mind that $\delta_A$
 and $\bar{\partial}_y$ do not commute. Indeed, a new $A$ appears each
time the
 operator $\bar{\partial}_y$ is applied on a quantity $Q$ since $\bar{\partial}_y
 Q=A
 \L_n Q$, where $\L_n=\partial_{\tilde d}$ is the
 Lie derivative along the normal direction. So $\delta_A Q =0$ whereas
$\delta_A \bar{\partial}_y Q=\hat{O}A'(y)$ where $\hat O$ is some
 four-dimensional operator. If we consider for instance the
 $y$-derivative of the extrinsic
curvature in
(\ref{Euv 1}), we have:
\ba
\delta_A \bar{\partial}_y K^{\mu}_{\, \nu}(y)&=&\delta_A K^{\mu\,
  \prime}_{\, \nu}(y)\notag \\
&=&\hspace{-3pt}-A'(y)E^{\mu}_{\nu}(y)-D^{\mu}D_{\nu}
A'(y)\label{K' A} \\
&&\hspace{-3pt}-A'(y)K^{\mu}_{\alpha}(y)K^{\alpha}_{\nu}(y)
+A'(y)\frac{1}{L^2}\delta^{\mu}_{\nu}.\notag
\ea
In the Taylor expansion, (\ref{taylor1}), there will a be a term,
which we denote by $\K^{\ \mu}_{1\, \nu}(0)$, that has only the first
derivative along $\bar{\partial}_y$ and all the other ones along $\delta_A$:
\ba
\K^{\ \mu}_{1\, \nu}(y)=\sum_{m\ge1}\frac{1}{m!}
 \delta_A^{(m-1)}\bar{\partial}_y K^{\mu}_{\, \nu}(y).
\ea
We may write
\ba
\delta_A^{(m-1)}\bar{\partial}_y K^{\mu}_{\, \nu}(y)
=\hat{O}^{\mu}_{\nu} A^{(m-1)},
\ea
where he operator $\hat{O}^{\mu}_{\nu}$ can be read off from (\ref{K' A}).
Using this relation, we have:
\ba
\K^{\ \mu}_{1\, \nu}(y)&=&\hat{O}^{\mu}_{\nu} \sum_{m\ge1}\frac{1}{m!}
A^{(m-1)}(y) \notag \\
&=&\hat{O}^{\mu}_{\nu}\int^{y+1}_{y}A(y')\ dy'.
\ea
In particular the term $\K^{\ \mu}_{1\, \nu}(0)$ that contributes to
the Taylor expansion (\ref{taylor1}) is:
\ba
\K^{\ \mu}_{1\, \nu}(0)&=&\hat{O}^{\mu}_{\nu}(y=0)\int_0^1A(y)\,dy \\
&=&\left. -d\, E^{\mu}_{\nu}-D^{\mu}D_{\nu}d\right|_{y=0},
\label{K1}
\ea
which is precisely the first term (\ref{K'(0)}) that was contributing in
the Taylor expansion when $A$ was assumed to be $y$-independent.

It what follow we shall see that the Taylor expansion can be expressed
as a sum of $\K_n$ which expression is precisely the same as $K^{(n)}$
when $A$ is assumed to be $y$-independent.

We shall work in a symbolic way, omitting any indices.
First we notice that the Taylor expansion (\ref{taylor1}, \ref{taylor2})
 can be written in the form:
\ba
\sum_{n\ge1}\frac{1}{n!}\K_n(0)=0,
\label{taylor4}
\ea
with
\ba
\frac{1}{n!}\K_n(y)&=&\hspace{-12pt}\sum_{m_1,\cdots, m_n\ge0}
\frac{1}{
l_n !
}\ \F_n^{m_1,\cdots, m_n},
\ea
where we wrote $l_n=(m_1+\cdots+m_n+n)$ and
\ba
\hspace{-20pt}\mathcal{F}_n^{m_1,\cdots, m_n}&=&\delta_A^{(m_n)}\bar{\partial}_y
\delta_A^{(m_{n-1})}\bar{\partial}_y
\cdots
\bar{\partial}_y\delta_A^{(m_1)}\bar{\partial}_y K.
\ea
Recalling that each time $K$ is differentiated with respect to $y$, a
new power of $A$ appears, so that
\ba
\bar{\partial}_y^{(n)} K =\hat{O}_n \left[A\hat{O}_{n-1}\left(
 \cdots \hat{O}_1 A\right)\right],
\ea
where
each
operator $\hat{O}_i$ depends on $y$ but includes only derivatives along the $y=\text{const}$
hypersurface and quantities $Q=K, E,q,\Gamma$.
Using this notation, $\F_n$ is:
\ba
&&\hspace{-20pt}
\F_n^{m_1,\cdots, m_n}=  \\
&&\hspace{-10pt}
\hat{O}_n\left[A
\hat{O}_{n-1}
\left(A\cdots
\hat{O}_2\left(
A \hat{O}_1 A^{(m_1)}\right)^{(m_2)}
 \right)^{(m_{n-1})}
\right]^{(m_n)},\notag
\ea
where the operators $\hat{O}_i$ should be treated as independent variable of $y$ so that for instance
\ba
\left(
A \hat{O}_1 A^{(m_1)}\right)^{(m_2)}=\sum C^k_{m_2}A^{(k)}\hat{O}_1
A^{(m_1+m_2-k)}.\hspace{5pt}
\ea
When all the sums are performed, we obtain:
\ba
&&\hspace{-30pt}
\frac{1}{n!}\K_n(y)= \label{Kn=d^n} \\
&&\hspace{-20pt}
\hat{O}_1
\int \left[
A \hat{O}_2
\left(
A
\int \cdots
\int \left(
A \hat{O}_n
\int
A dy
\right)dy
\right)dy
\right]dy. \notag
\ea
So on the brane, we can express $\K_n$ in the form:
\ba
\K_n(0)&=&n!\, U_1(1) \\
U_{m}(y)&=&\hat{O}_m \int_0^y A(y)\, U_{m+1}(y) dy,\ \ \ m < n\hspace{10pt}\\
U_n(y)&=&\hat{O}_n \int_0^y A(y) dy=\hat{O}_n \tilde{d}(y).
\ea
So far this result is exact. When the branes are close, this result will simplify remarkably.
First we point out that $0 <\tilde{d}(y) < d \ll L$ for any $0<y<1$. This follows from the fact that the
bulk geometry is completely regular, so between the branes, $A(y,x) > 0$, which implies
$0<\int_0^y A(y,x) dy < d(x)\ll L$ for any $0<y<1$. This holds similarly for any multiple integral.
The same simplifications as in the previous Section will therefore be valid here. For instance if in the
previous Section we had $\hat{O}_1 d \simeq D d$, then this will remain true in this case as well:
$\hat{O}_1 \int_0^y A(y,x) dy \simeq D \int_0^y A(y,x)$.

In the following we shall label with a ``bar'' any quantity which was derived in the previous Section assuming
$A$ was independent of $y$.
In the previous Section, we had $\bar A=d$ and so $U_m$ was simply:
\ba
\bar{U}_m(y)=\frac{y^{n+1-m}}{(n+1-m)!}\hat{O}_m\left[d \hat{O}_{m+1}\left(
d \cdots \hat{O}_n d \right)\right],\hspace{10pt}
\ea
so we had:
\ba
\bar{K}^{(n)}(0)&=&
\hat{O}_1 \left[
d \, \hat{O}_2\left(d \cdots d\,  \hat{O}_1d \right)
\right],
\ea
using the fact that at $d \ll L$,
the action of the operators $\hat{O}_i$ was considerably simplified
and at leading order in
$d/L$, they were equivalent to the overall operator $\hat{O}^{(n)}$:
\ba
\bar{K}^{(n)}(0)&=&\hat{O}^{(n)} d^n \label{simpli1}\\
\text{with} \
\hat{O}^{(2n+1)}&=&\frac{1}{(2n+1)!}\left(-E(0)-D^2\right)D^{(2n)}
\hspace{25pt} \label{simpli2}\\
\hat{O}^{(2n)}&=&\frac{1}{2n!}K(0)D^{(2n)}, \label{simpli3}
\ea
where $D$ is a derivative along the $y=\text{const}$
hypersurface. The leading contribution was indeed:
\ba
\bar{K}^{(2n+1)}(0)&=&K'(0)\ \left(\partial d\right)^{2n}\\
\bar{K}^{(2n)}(0)&=&K(0)\ \left(\partial d\right)^{2n}.
\ea
Now we may go back to the situation where $A$ has some $y$-dependence. Because
$0<\int_0^y A(y,x) dy < d(x)\ll L$ for any $0<y<1$ and similarly for any multiple integral in $U_m$,
when the operators $\hat{O}_i$ are applied on these integrals we can reproduce step by step exactly the same
procedure as we followed in the previous Section in order to keep only
the leading order in $d/L$. In particular, the repeated action of each
$\hat{O}_i$ on each multiple integral can be substituted, in the
small $d$ limit (which implies a small $\int_0^y \cdots \int_0^y A dy$
limit) by the action of
an overall operator $\hat{O}^{(n)}$.
If we do so, the leading contribution can be expressed in the
same way as in (\ref{simpli1}):
\ba
\frac{1}{n!}\K_{n}(0)&=& \hat{O}^{(n)} \left[
\int_0^1 A
\left(
\int_0^y \cdots
A \int_0^y
A dy
\right)
\right],\hspace{20pt}
\ea
with the operator $\hat{O}^{(n)}$ as given in (\ref{simpli2},\ref{simpli3}).
Now the multiple integral is simply:
\ba
\int_0^1 A
\left(
\int_0^y \cdots
A \int_0^y
A dy
\right)&=&\frac{1}{n!}\left(\int_0^1 A dy\right)^n \notag\\
&=&\frac{d^n}{n!}.
\ea
So the rest follows exactly as in the previous case. For any $n$, the leading contribution to $\K_n$ is:
\ba
\K_{n}(0)&=&\hat{O}^{(n)}d^n,
\ea
exactly as in (\ref{simpli1}).
 So the leading order in $d/L$ of the Taylor
 expansion (\ref{taylor4}) is independent of the $y$-dependence of $A$,
 and in particular we obtain the same result whether $A=d$ or any other
 function of $y$ such that $\int_0^1A\, dy=d$.
 The close brane limit
 theory derived in Section \ref{closebranes} is therefore valid independently of the precise
 expression of $A$, as this should be since our result is gauge
 invariant.

%
%
\section{Divergence of the stress-energy tensor}
\label{appendix div T}

In order to work out in detail the divergence of the stress-energy
tensor, we will consider directly the situation with a potential in
section \ref{section small d with V}.
We therefore consider the stress-energy tensor as given in (\ref{Tuv + V}) with the
equation of motion for $d$ given in (\ref{box d +V}). We can work
in the situation where no potential is present as in section
(\ref{section Conservation of Energy}) by setting $V=W=0$ in the
following.

Using (\ref{Tuv + V}), the divergence of the stress-energy tensor may
be written as:
\ba
d\, D_\mu T^\mu _\nu&=&-V'(d) \partial_\nu d
-T^\mu_\nu\, \partial_\mu d
+R^\mu_{\,\nu} \partial_\mu d \label{D Tuv 3}\\ &&\hspace{-50pt}
 -\frac{1}{L}\left(6z^2 f'+4 z^4 g'+2g\right) \partial_\nu
z^2
+\frac{2}{L}z^2g'F^\mu_{\, \nu}\partial_\mu z^2 \notag \\
&&\hspace{-50pt}
+\frac{1}{L}g D_\mu F^\mu_{\, \nu}
-\partial_\mu d \left[\bar{E}^\mu_\nu-\bar E \delta^\mu_\nu \right]
- d D_\mu\left[\bar{E}^\mu_\nu-\bar E\,  \delta^\mu_\nu \right],\notag
\ea
where again $f$ and $g$ are taken as function of $z^4$.
We stress that $\bar{E}^\mu_\nu \sim d^0$ so $ d D_\mu \bar{E}^\mu_\nu $ will
 be neglected in what follows.\\
The first line simplifies, using the relation
\ba
T^\mu _\nu&=&R^\mu_{\,\nu}+\frac{1}{2}T\delta^{\mu\nu}\nn\\
&=&R^\mu_\nu-\frac{1}{2d}\left(3W+4V \right)\delta^\mu_\nu.
\ea
>From the Einstein equation and the equation for $d$ (\ref{box d +V}), we have:
\ba
\partial_\mu z^2&=&-2 D_\mu D_\nu d \, \partial^\nu d \notag \\
&=& -2\left(-\frac{1}{L}(f g^{(+)}_{\mu \nu}+g\, F_{\mu
  \nu})+(V+W)g^{(+)}_{\mu \nu}\right)\partial^\nu d \notag \\
&=&\frac{2}{L}(f-z^2 g)\partial_\mu d
-2(V+W)\partial_\mu d,
\ea
where terms of order $d/L^2$ have been omitted.\\
Using this relation and
\ba
D_\mu  F^\mu_{\, \nu}=
2\Box d\,  \partial_\nu d
=-\frac{2}{L}
\left(
2 z^2 g+4 f-LW
 \right)\partial_\nu d,
\ea
the expression (\ref{D Tuv 3}) simplifies to:
\ba
 d\, D_\mu T^\mu_\nu&=&
- \left[\bar E^\mu _{\nu}-\bar E \delta^\mu_\nu
 \right]\partial_\mu d
 \label{Du Tuv 4} \\
&&- \frac{12}{L^2}\left[g f+z^2\left(f-z^2 g\right)\left(f'+z^2
 g'\right)\right]\partial_\nu d\notag \\
&&+\Big[
-V'(d)+\frac{3W+4V}{2d}
+\frac{2g}{L}\left(3W+2V \right) \notag\\
&& \phantom{+\Big[} +\frac{12}{L}z^2\left(f'+z^2 g'\right)\left(V+W
\right)
\Big]\partial_\nu d \notag\\
&& + \frac{1}{L^2}\mathcal{O} \left(\frac{d}{L}\right).\notag
\ea
In the absence of a potential, the divergence of $T^\mu_\nu$
 simplifies to the first two lines. Both
\ba
\hspace{-20pt}\left[\bar E^\mu _{\nu}-\bar E \delta^\mu_\nu
 \right]\partial_\mu d&=&\hspace{-5pt}-
\frac{12}{L^2}\left[g f \right. \\
&&\hspace{-20pt}\left.+z^2\left(f-z^2 g\right)\left(f'+z^2
 g'\right)\right]\partial_\nu d\nn
\ea
and, in the presence of a potential,
\ba
V'(d)&=&\frac{3W+4V}{2d}
+\frac{2g}{L}\left(3W+2V \right)\\
&&  +\frac{12}{L}z^2\left(f'+z^2 g'\right)\left(V+W
\right) \notag
\ea
need to hold independently, in the close brane limit.
%
%
\section{Evolution of the Weyl Tensor}
\label{appendix weyl}
In this Section we derive the evolution equation (\ref{E prime}) for
$E^\mu_\nu$. Essentially the same results are quoted in
\cite{Shiromizu:2002qr,Shiromizu:2002ve,Aliev:2004ds}
but we repeat them here for completeness. We assume the general form
(\ref{ds^2=})
for the
metric.
$n=A^{-1}\partial/\partial y$ is the unit normal to
hypersurfaces of constant $y$ and, for a purely four-dimensional
tensor, the Lie derivative with respect to $n$ is given by
\[
\L_n=n=A^{-1}\partial/\partial y
\]
The projection tensor $h_{ab}=g_{ab}-n_a n_b$
then satisfies
$h^5_5=h^\mu_5=h^5_\mu=0,\ h^\mu_\nu=\delta^\mu_\nu$
and so can be left implicit in the index convention. The
four-dimensional and five-dimensional Riemann tensors are related by
\be
\label{w1}
R_{\mu\nu\alpha\beta}=\rfive_{\mu\nu\alpha\beta}+2K_{\mu[\alpha}K_{\beta]\nu}\ee
The electric and magnetic parts of the bulk
Riemann tensor are defined to be
\be
\label{w2}
\tilde{E}_{ab}\equiv \rfive_{acbd}n^c n^d\qquad \tilde{B}_{abc}\equiv h_a^e h_b^f n^d
\ \rfive_{efcd}\ee
and are purely four-dimensional. (\ref{Euv 1}) then follows identically
from a direct evaluation of the Riemann tensor for the metric
(\ref{ds^2=}). Denoting by $\nabla_a$ the five-dimensional covariant
derivative, we find that
\ba
\label{w3}
\frac{1}{A}\del_5 \left(\rfive_{\mu\nu\alpha c}n^c\right)&=&\mathcal{L}_n\Bt_{\mu\nu\alpha}+\frac{2}{A}\partial_{[\mu}A\Et_{\nu]\alpha}\notag\\&&+2K^\beta_{[\mu}\Bt_{\nu]\beta\alpha}-K^\beta_\alpha\Bt_{\mu\nu\beta}\\
\label{w4}
\frac{1}{A}\del_\mu\left(\rfive_{\nu 5\alpha c}
  n^c\right)&=&D_\mu\Et_{\alpha\nu}-K^\beta_\mu\Bt_{\nu\beta\alpha}\ea
As a consequence of the five-dimensional Bianchi identities, we
  find
\[
\del_{[5}\left(\rfive_{\mu\nu]\alpha c}
  n^c\right)\equiv\left(\del_{[5}n^c\right)\rfive_{\mu\nu]\alpha c}\]
which, from (\ref{w3}) and (\ref{w4}), gives
\ba
\L_n\Bt_{\mu\nu\alpha}&=&
K^\beta_\alpha\Bt_{\mu\nu\alpha}-\frac{2}{A}\partial_{[\mu}A\Et_{\nu]\alpha}-2D_{[\mu}\Et_{\nu]\alpha}\nn\\
\label{w5}
&&-2\Bt_{\alpha\beta[\mu}K^\beta_{\nu]}-\frac{\partial^\beta A}{A}\ \rfive_{\mu\nu\alpha\beta}\ea
Similarly $\del_{[5}\rfive_{\mu\nu]\alpha\beta}\equiv0$ yields
\ba
0&=&n\cdot\partial
\rfive_{\mu\nu\alpha\beta}-\frac{2}{A}\Bt_{\mu\nu[\alpha}\partial_{\beta]}A-\frac{2}{A}\Bt_{\alpha\beta[\mu}\partial_{\nu]}A\nn\\
&&+2\rfive_{\mu\nu\rho[\alpha}K^\rho_{\beta]}+2D_{[\mu}\Bt_{|\alpha\beta|\nu]}+2K_{\mu[\beta}\Et_{\alpha]\nu}\nn\\
&&-2K_{\nu[\beta}\Et_{\alpha]\mu}\nn\ea
(\ref{w1}) and (\ref{Euv 1}) can then be used to write this purely in
terms of four-dimensional quantities, giving
\ba
\label{w6}
\L_n R_{\mu\nu\alpha\beta}&=&-2 R_{\mu\nu\rho[\alpha}
  K^\rho_{\beta]}+\frac{2}{A}\Bt_{\mu\nu[\alpha}\partial_{\beta]}A\\
&&-2D_{[\mu}\Bt_{|\alpha\beta|\nu]}+\frac{2}{A}\Bt_{\alpha\beta[\mu}\partial_{\nu]A}\nn\\
&&-\frac{2}{A}D_\mu\partial_{[\alpha}A K_{\beta]\nu}
 +\frac{2}{A}D_\nu\partial_{[\alpha}A K_{\beta]\mu}\nn\ea
All the above is true identically off-shell, i.e. not using the bulk Einstein equations $\rfive_{ab}=-\frac{4}{L^2}g_{ab}$
These relate the projections of the Riemann tensor to those of the Weyl
tensor,
\ba
E_{\mu\nu}&\equiv& ^{(5)}C_{\mu c \nu d}n^c
n^d=\Et_{\mu\nu}+\frac{q_{\mu\nu}}{L^2}\nn\\
B_{\mu\nu\sigma}&\equiv&^{(5)}C_{\mu\nu\sigma c}n^c=\Bt_{\mu\nu\sigma}\nn\ea
and we can use (\ref{gauss codacci}) to write
\ba
\L_n E_{\mu\nu}&=&\L_n
f_{\mu\nu}-\L_n\ R_{\mu\nu}-\frac{6}{L^2}K_{\mu\nu}\nn\\
f_{\mu\nu}&\equiv&KK_{\mu\nu}-K_{\mu\rho}K^\rho_\nu\nn\ea
The first term on the RHS can be found with repeated applications of
(\ref{Euv 1}) whilst the second can be found from (\ref{w6}); the
result is
\ba
\L_n E_{\mu\nu}&=&D^\alpha
B_{\alpha(\mu\nu)}+\frac{1}{L^2}\left(Kq_{\mu\nu}-K_{\mu\nu}\right)\nn\\
&&+K^{\alpha\beta}\ R_{\mu\alpha\nu\beta}+3K^\alpha_{(\mu}E_{\nu)\alpha}-KE_{\mu\nu}\nn\\
&&+K^{\alpha\beta}\left(K_{\alpha\mu}K_{\beta\nu}-K_{\alpha\beta}K_{\mu\nu}\right)+\frac{2}{A}\partial^\alpha
AB_{\alpha(\mu\nu)}\nn\ea
This can be written in terms of quantities which vanish on the brane
where $K_{\mu\nu}\propto q_{\mu\nu}$, namely the traceless part
$\hat{K}^\mu_\nu\equiv K^\mu_\nu-\frac{1}{4}K\delta^\mu_\nu$ of the
extrinsic curvature and the Weyl tensor (which will vanish when
contracted with $K^{\alpha\beta}$):
\ba
\label{w7}
\L_n
E_{\mu\nu}&=&2K^\alpha_\nu
E_{\mu\alpha}-\frac{3}{2}KE_{\mu\nu}-\frac{1}{2}K^\alpha_\beta
E^\beta_\alpha q_{\mu\nu}\\
&&+C_{\mu\alpha\nu\beta}K^{\alpha\beta}+2\hat{K}^\alpha_\mu
\hat{K}_{\alpha\beta}
\hat{K}^\alpha_\nu-\frac{7}{6}\hat{K}_{\alpha\beta}\hat{K}^{\alpha\beta}\hat{K}_{\mu\nu}\nn\\
&&-\frac{1}{2}q_{\mu\nu}\hat{K}_{\alpha\beta}\hat{K}^\beta_\rho\hat{K}^{\alpha\rho}
+D^\alpha B_{\alpha(\mu\nu)}\nn\ea
which reduces to (\ref{E prime}) using
$B_{\mu\nu\rho}=2D_{[\mu}K_{\nu]\rho}$. \vspace{-30pt}
%
%
%

\end{document}